\documentclass{article}

\usepackage{arxiv}

\usepackage[utf8]{inputenc} 
\usepackage[T1]{fontenc}    
\usepackage{hyperref}       
\usepackage{url}            
\usepackage{booktabs}       
\usepackage{amsfonts}       
\usepackage{nicefrac}       
\usepackage{microtype}      
\usepackage{lipsum}		
\usepackage{graphicx}
\usepackage{natbib}
\usepackage{doi}
\usepackage{color}
\usepackage{amsmath}
\usepackage{quantikz}
\usepackage{cite}
\usepackage{braket}
\usepackage{xcolor}
\usepackage{algorithm}
\usepackage{algpseudocode}
\newcommand{\gb}{{\vec{\beta}, \vec{\gamma}}}

\title{Similarity-Based Parameter Transferability in the Quantum Approximate Optimization Algorithm}

\author{ {\hspace{1mm}Alexey Galda} \\
	James Franck Institute \\
	University of Chicago \\
	Chicago, IL 60637 \\
	\texttt{alex.galda@gmail.com} \\
	\And
	{\hspace{1mm}Eesh Gupta} \\
	Department of Physics and Astronomy\\
	Rutgers University\\
	Piscataway, NJ 08854 \\
    \And
    {\hspace{1mm}Jose Falla} \\
    Department of Physics and Astronomy \\
    University of Delaware \\
    Newark, DE 19716 \\
    \And
    {\hspace{1mm}Xiaoyuan Liu} \\
    Fujitsu Research of America, Inc. \\
    Sunnyvale, CA 94085 \\
    \And
    {\hspace{1mm}Danylo Lykov} \\
    Computational Science Division \\
    Argonne National Laboratory \\
    Lemont, IL 60439 \\
    \And
    {\hspace{1mm}Yuri Alexeev} \\
    Computational Science Division \\
    Argonne National Laboratory \\
    Lemont, IL 60439 \\
    \And
    {\hspace{1mm}Ilya Safro} \\
    Department of Computer and Information Sciences \\
    University of Delaware \\
    Newark, DE 19716 \\
}

\hypersetup{
pdftitle={Similarity-Based Parameter Transferability in the Quantum Approximate Optimization Algorithm},
pdfauthor={Alexey Galda, Eesh Gupta, Jose Falla, Xiaoyuan Liu, Danylo Lykov, Yuri Alexeev, Ilya Safro},
pdfkeywords={Quantum Computing, Quantum Approximate Optimization Algorithm, Quantum Optimization, Quantum Software, Parameter Transferability},
}

\begin{document}
\maketitle

\begin{abstract}
The quantum approximate optimization algorithm (QAOA) is one of the most promising candidates for achieving quantum advantage through quantum-enhanced combinatorial optimization. A near-optimal solution to the combinatorial optimization problem is achieved by preparing a quantum state through the optimization of quantum circuit parameters. Optimal QAOA parameter concentration effects for special MaxCut problem instances have been observed, but a rigorous study of the subject is still lacking. In this work we show clustering of optimal QAOA parameters around specific values;  consequently, successful transferability of parameters between different QAOA instances can be explained and predicted based on local properties of the graphs, including the type of subgraphs (lightcones) from which graphs are composed as well as the overall degree of nodes in the graph (parity). We apply this approach to several instances of random graphs with a varying number of nodes as well as parity and show that one can use optimal donor graph QAOA parameters as near-optimal parameters for larger acceptor graphs with comparable approximation ratios. This work presents a pathway to identifying classes of combinatorial optimization instances for which variational quantum algorithms such as QAOA can be substantially accelerated.
\end{abstract}


\section{Introduction}
Quantum computing seeks to exploit the quantum mechanical concepts of entanglement and superposition to perform a computation that is significantly faster and more efficient than what can be achieved by using the most powerful supercomputers available today~\citep{preskill_quantum_2018, arute_quantum_2019}. Demonstrating quantum advantage with optimization algorithms~\citep{Alexeev2021} is poised to have a broad impact on science and humanity by allowing us to solve problems on a global scale, including finance~\citep{herman2022survey}, biology~\citep{outeiral2021prospects}, and energy~\citep{joseph2023quantum}. Variational quantum algorithms, a class of hybrid quantum-classical algorithms, are considered primary candidates for such tasks and consist of parameterized quantum circuits with parameters updated in classical computation. The quantum approximate optimization algorithm (QAOA)~\citep{Hogg2000quantumsearch, Hogg2000, farhi2014quantum, hadfield2017quantum} is a variational algorithm for solving classical combinatorial optimization problems. In the domain of optimization on graphs, it is most often used to solve NP-hard problems such as MaxCut~\citep{farhi2014quantum}, community detection~\citep{shaydulin2019network}, and partitioning~\citep{ushijima2021multilevel} by mapping them onto a classical spin-glass model (also known as the Ising model) and minimizing the corresponding energy, a task that in itself is NP-hard.

In this work we demonstrate two related key elements of optimal QAOA parameter transferability. First, by analyzing the distributions of subgraphs from two QAOA MaxCut instance graphs, one can predict how close the optimized QAOA parameters for one instance are to the optimal QAOA parameters for another. Second, by analyzing the overall parity of both donor-acceptor pairs, one can predict good transferability between those QAOA MaxCut instances. The measure of transferability of optimized parameters between MaxCut QAOA instances on two graphs can be expressed through the value of the approximation ratio, which is defined as the ratio of the energy of the corresponding QAOA circuit, evaluated with the optimized parameters $\gamma, \beta$, divided by the energy of the optimal MaxCut solution for the graph.

While the optimal solution is not known in general for relatively small instances (graphs with up to 256 nodes are considered in this paper), it can be found by using classical algorithms, such as the Gurobi solver \citep{gurobi}\footnote{The Gurobi solver provides  classically optimal MaxCut solutions in a competitive speed with known optimization gap. For the purpose of this work, there is no particular reason to choose Gurobi over IPOPT or other similarly performing solvers.}. We first focus our attention on similarity based on the subgraph decomposition of random graphs and show that good transferability of optimized parameters between two graphs is directly determined by the transferability between all possible permutations of pairs of individual subgraphs. The relevant subgraphs of these graphs are defined by the QAOA quantum circuit depth parameter $p$. In this work we focus on the case $p = 1$; however, our approach can be extended to larger values of $p$. Higher values of $p$ lead to an increasing number of subgraphs to be considered, but the general idea of the approach remains the same. This question is beyond the scope of this paper and will be addressed in our future work. We then move to similarity based on graph parity and determine that we can predict good optimal parameter transferability between donor-acceptor graph pairs with similar parities. Here, too,  more work remains to be done regarding the structural effects of graphs on optimal parameter transferability.

Based on the analysis of the mutual transferability of optimized QAOA parameters between all relevant subgraphs for computing the MaxCut cost function of random graphs, we show good transferability \emph{within} the classes of odd and even random graphs of arbitrary size. We also show that transferability is poor \emph{between} the classes of even and odd random graphs, in both directions, based on the poor transferability of the optimized QAOA parameters between the subgraphs of the corresponding graphs. When considering the most general case of arbitrary random graphs, we construct the transferability map between all possible subgraphs of such graphs, with an upper limit of node connectivity ${d_\mathrm{max} = 6}$, and use it to demonstrate that in order to find optimized parameters for a MaxCut QAOA instance on a large 64-, 128-, or 256-node random graph, under specific conditions, one can reuse the optimized parameters from a random graph of a much smaller size, ${N = 6}$, with only a $\sim$1\% reduction in the approximation ratio.

This paper is structured as follows. In Section~\ref{sec:QAOA} we present the relevant background material on QAOA. In Section~\ref{sec:transfer} we consider optimized QAOA parameter transferability properties between all possible subgraphs of random graphs of degree up to ${d_\mathrm{max} = 6}$. We then extend the consideration to parameter transferability using graph parity as a metric, and we demonstrate the power of the proposed approach by performing optimal transferability of QAOA parameters in many instances of donor-acceptor graph pairs of differing sizes and parity. We find that one can effectively transfer optimal parameters from smaller donor graphs to larger acceptor graphs, using similarities based on subgraph decomposition and parity as indicators of good transferability. In Section~\ref{sec:conclusions} we conclude with a summary of our results and an outlook on future advances with our approach.

\section{QAOA}\label{sec:QAOA}

The quantum approximate optimization algorithm is a hybrid quantum-classical algorithm that combines a parameterized quantum evolution with a classical outer-loop optimizer to approximately solve binary optimization problems~\citep{farhi2014quantum,hadfield2019quantum}. QAOA consists of $p$ layers (also known as the circuit depth) of pairs of alternating operators, with each additional layer increasing the quality of the solution, assuming perfect noiseless execution of the corresponding quantum circuit. With quantum error correction not currently supported by modern quantum processors, practical implementations of QAOA are limited to ${p \leq 3}$ because of noise and limited coherence of quantum devices imposing strict limitations on the circuit depth~\citep{zhou_quantum_2020}. Motivated by the practical relevance of results, we focus on the case ${p = 1}$ in this paper.

\subsection{QAOA Background}\label{subsec:theory}

Consider a combinatorial problem defined on a space of binary strings of length $N$ that has $m$ clauses. Each clause is a constraint satisfied by some assignment of the bit string. The objective function can be written as ${C(z) = \sum_{\alpha=1}^m C_{\alpha}(z)}$, where $z = z_1z_2\cdots z_N$ is the bit string and ${C_{\alpha}(z) = 1}$ if $z$ satisfies the clause $\alpha$, and 0 otherwise. QAOA maps the combinatorial optimization problem onto a $2^N$-dimensional Hilbert space with computational basis vectors $\ket{z}$ and encodes $C(z)$ as an operator $C$ diagonal in the computational basis.

At each call to the quantum computer, a trial state is prepared by applying a sequence of alternating quantum operators 

\begin{equation} \label{eq:QAOAstate}
    \ket{\gb}_p := U_B(\beta_p)U_C(\gamma_p)\ldots U_B(\beta_1)U_C(\gamma_1)\ket{s}\,,
\end{equation}
where $U_C(\gamma) = e^{-i\gamma C}$ is the phase operator; $U_B(\beta)=e^{-i\beta B}$ is the mixing operator, with $B$ defined as the operator of all single-bit $\sigma^x$ operators; $B = \sum_{j=1}^N \sigma_j^x$; and $\ket{s}$ is some easy-to-prepare initial state, usually taken to be the uniform superposition product state. The parameterized quantum circuit (\ref{eq:QAOAstate}) is called the QAOA \emph{ansatz}. We refer to the number of alternating operator pairs~$p$ as the QAOA \textit{depth}. The selected parameters $\vec{\beta},\vec{\gamma}$ are said to define a \emph{schedule}, analogous to a similar choice in quantum annealing.

\begin{figure*}[h!]
    \centering
    \includegraphics[width=\linewidth]{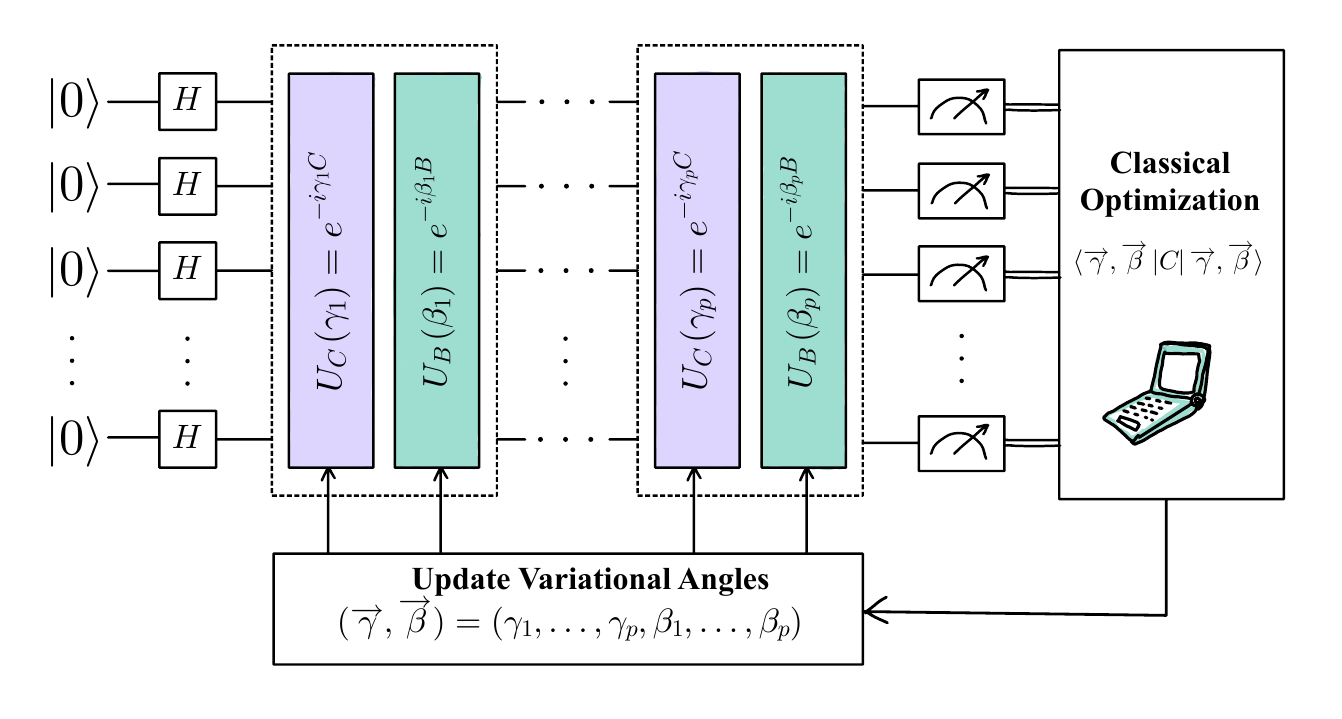}
    \caption{Schematic pipeline of a QAOA circuit. A parametrized ansatz is initialized, followed by series of applied unitaries that define the depth of the circuit. Finally, measurements are made in the computational basis, and the variational angles are classically optimized. This hybrid quantum-classical loop continues until convergence to an approximate solution is achieved.}
    \label{fig:qaoa_schem}
\end{figure*}

Preparation of the state (\ref{eq:QAOAstate}) is followed by a measurement in the computational basis. The output of repeated state preparation and measurement may be used by a classical outer-loop algorithm to select the schedule~$\vec{\beta},\vec{\gamma}$. We consider optimizing the expectation value of the objective function

\[\langle C\rangle_p = \bra{\vec{\beta}, \vec{\gamma}}_pC\ket{\vec{\beta}, \vec{\gamma}}_p \,,
\]
as originally proposed in~\citep{farhi2014quantum}. The output of the overall procedure is the best bit string $z$ found for the given combinatorial optimization problem. Figure \ref{fig:qaoa_schem} presents a schematic pipeline of the QAOA algorithm. We emphasize that the task of finding good QAOA parameters is challenging in general, for example because of encountering  barren plateaus~\citep{anschuetz_beyond_2022, wang_noise-induced_2021}. Acceleration of the optimal parameters search for a given QAOA depth $p$ is the focus of many approaches aimed at demonstrating  quantum advantage. Examples include  warm- and multistart  optimization~\citep{egger2020warmstarting,shaydulin2019multistart}, problem decomposition~\citep{shaydulin2019hybrid},  instance structure analysis~\citep{shaydulin2020classical}, and parameter learning~\citep{khairy2020learning}.

\subsection{MaxCut}\label{subsec:maxcut}

For  studying the transferability of optimized QAOA parameters, we consider the MaxCut combinatorial optimization problem. Given an unweighted undirected simple graph ${G = (V,E)}$, the goal of the MaxCut problem is to find a partition of the graph's vertices into two complementary sets such that the number of edges between the two sets is maximized. In order to encode the problem in the QAOA setting, the input is a graph with ${|V| = N}$ vertices and ${|E| = m}$ edges, and the goal is to find a bit string $z$ that maximizes \begin{eqnarray} C = \sum_{jk\in E} C_{jk}, \label{eq:maxcut_cost} \end{eqnarray}
where \begin{eqnarray*}
C_{jk} = \frac{1}{2}(-\sigma_j^z \sigma_k^z + 1).\end{eqnarray*}

It has been shown in \citep{farhi2014quantum} that on a 3-regular graph, QAOA with $p=1$ produces a solution with an approximation ratio of at least 0.6924.

\subsection{QAOA Simulator and Classical MaxCut Solver}

Calculating the approximation ratio for a particular MaxCut problem instance requires the optimal solution of the combinatorial optimization problem. This problem is known to be NP-hard, and classical solvers require exponential time to converge. For our experiments, we use the Gurobi solver \citep{gurobi} with the default configuration parameters, running the solver until it converges to the optimal solution. For our QAOA simulations, we use QTensor \citep{qtensor}, a large-scale quantum circuit simulator with step-dependent parallelization. QTensor simulates circuits based on a tensor network approach, and as such, it can provide an efficient approximation to certain classes of quantum states~\citep{kardashin_quantum_2021, biamonte_tensor_2017}.

\section{Parameter Transferability}\label{sec:transfer}

Solving a QAOA instance calls for two types of executions of quantum circuits: iterative optimization of the QAOA parameters and the final sampling from the output state prepared with those parameters. While the latter is known to be impossible to simulate efficiently for large enough instances using classical hardware instead of a quantum processor~\citep{farhi2014quantum}, the iterative energy calculation for the QAOA circuit during the classical optimization loop can be efficiently performed by using tensor network simulators for instances of a wide range of sizes~\citep{lykov2020tensor}, as described in the preceding section. This is achieved by implementing considerable simplifications in how the expectation value of the problem Hamiltonian is calculated by employing a mathematical reformulation based on the notion of the reverse causal cone introduced in the seminal QAOA paper~\citep{farhi2014quantum}. Moreover, in some instances, the entire search of the optimal parameters for a particular QAOA instance can be circumvented by reusing the optimized parameters from a different ``related'' instance, for example for which the optimal parameters are concentrated in the same region.

Optimizing QAOA parameters for a relatively small graph, called the donor, and using them to prepare the QAOA state that maximizes the expectation value $\langle C\rangle_p$ for the same problem on a larger graph, called the acceptor, is what we define as \emph{successful optimal parameter transferability}, or just transferability of parameters, for brevity. The transferred parameters can be used either directly without change, as implemented in this paper, or as a ``warm start'' for further optimization. In either case, the high computational cost of optimizing the QAOA parameters, which grows rapidly as the QAOA depth $p$ and the problem size are increased, can be significantly reduced. This approach presents a new direction for dramatically reducing the overall runtime of QAOA.

Optimal QAOA parameter concentration effects have been reported  for several special cases, mainly focusing on random 3-regular graphs~\citep{brandao2018fixed, streif2020training, akshay2021parameter}. Brandao et al.~\citep{brandao2018fixed} observed that the optimized QAOA parameters for the MaxCut problem obtained for a 3-regular graph are also nearly optimal for all other 3-regular graphs. In particular, the authors noted that in the limit of large $N$, where $N$ is the number of nodes, the fraction of tree graphs asymptotically approaches 1. We note that, for example, in the sparse Erd\"os--R\'enyi graphs, the trees are observed in short-distance neighborhoods with very high probability~\citep{newman2018networks}. As a result, in this limit, the objective function is the same for all 3-regular graphs, up to order $1/N$. 

The central question of this manuscript is  under what conditions the optimized QAOA parameters for one graph also maximize the QAOA objective function for another graph. To answer that question, we study transferability between subgraphs of a graph, since the QAOA objective function is fully determined by the corresponding subgraphs of the instance graph, as well as transferability between graphs of similar parities, in order to determine structural effects of graphs on effective transferability.

\subsection{Subgraph Transferability Analysis}\label{subsec:subgraphs}

\begin{figure*}[h!]
    \centering
    \includegraphics[width=\linewidth]{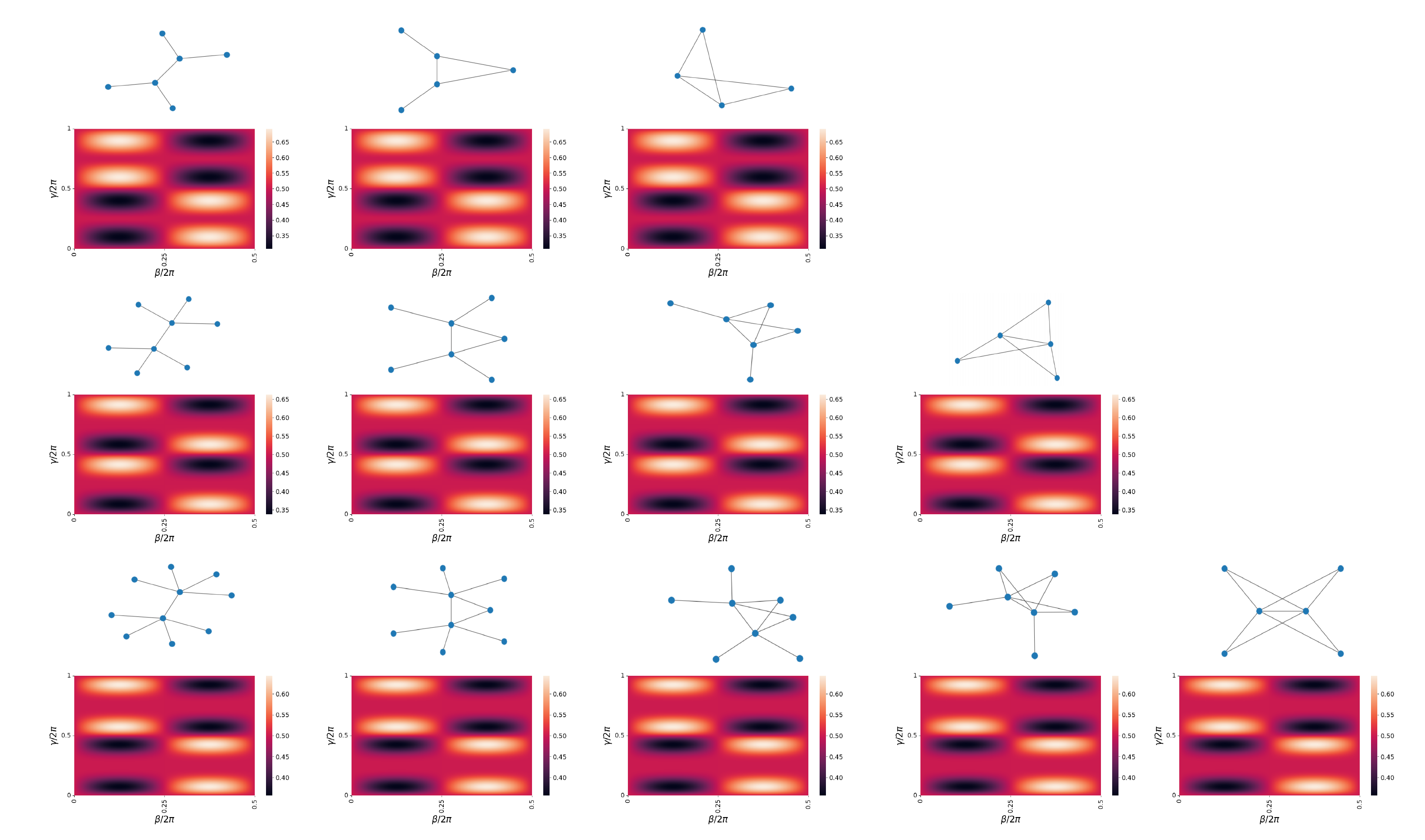}
    \caption{Landscapes of energy contributions for individual subgraphs of 3- (top row), 4- (middle row), and 5-regular (bottom row) random graphs, as a function of QAOA parameters $\beta \in ({0, \pi})$ and $\gamma \in (0, 2\pi)$. All subgraphs of 3- and 5- regular graphs have maxima located in the relative vicinity of one another. Subgraphs of 4-regular graphs also have closely positioned maxima between themselves; however, only half of them match with the maxima of subgraphs of odd-regular random graphs.}
    \label{fig:landscapes}
\end{figure*}

It was shown in the seminal QAOA paper~\citep{farhi2014quantum} that the expectation value of the QAOA objective function, $\langle C\rangle_p$, can be evaluated as a sum over contributions from subgraphs of the original graph, provided its degree is bounded and the diameter is larger than $2p$ (otherwise, the subgraphs cover the entire graph itself). The contributing subgraphs can be constructed by iterating over all edges of the original graph and selecting only the nodes that are $p$ edges away from the edge. Through this process, any graph can be deconstructed into a set of subgraphs for a given $p$, and only those subgraphs contribute to $\langle C\rangle_p$, as also discussed in Section~\ref{sec:QAOA}.

We begin by analyzing the case of MaxCut instances on 3-regular random  graphs for QAOA circuit depth $p = 1$, which have three possible subgraphs~\citep{farhi2014quantum, brandao2018fixed}. Figure \ref{fig:landscapes} (top row) shows the landscapes of energy contributions from these subgraphs, evaluated for a range of $\gamma$ and $\beta$ parameters. We can see that all maxima are located in the approximate vicinity of each other. As a result, the parameters optimized for any of the three graphs will also be near-optimal for the other two. Because any random 3-regular graph can be decomposed into these three subgraphs, for QAOA with $p = 1$, this guarantees that optimized QAOA parameters can be successfully transferred between any two 3-regular random graphs, which is in full agreement with~\citep{brandao2018fixed}.

The same effect is observed for subgraphs of 4-regular; see Figure~\ref{fig:landscapes} (middle row). The optimized parameters are mutually transferable between all four possible subgraphs of 4-regular graphs. Notice, however, that the locations of exactly half of all maxima for the subgraphs of 4-regular graphs do not match with those for 3-regular graphs. This means that one cannot expect good transferability of optimized parameters across MaxCut QAOA instances for 3- and 4-regular random graphs if these optimal parameters are to be transferred directly. It has been recently shown in \citep{basso_et_al:LIPIcs.TQC.2022.7} that gamma parameters can be rescaled in order to generalize between different random d-regular graphs.

Focusing now on all five possible subgraphs of 5-regular graphs, Figure~\ref{fig:landscapes} (bottom row), we notice that, again, good parameter transferability is expected between all instances of 5-regular random graphs. Moreover, the locations of the maxima match well with those for 3-regular graphs, indicating good transferability across 3- and 5-regular random graphs.

We discuss parameter concentration for instances of random graphs in a later section;  similar discussions can be found in \citep{brandao2018fixed} and \citep{wurtz2021fixed} in the context of 3-regular graphs.

\begin{figure*}[h!]
    \centering
    \includegraphics[width=\columnwidth]{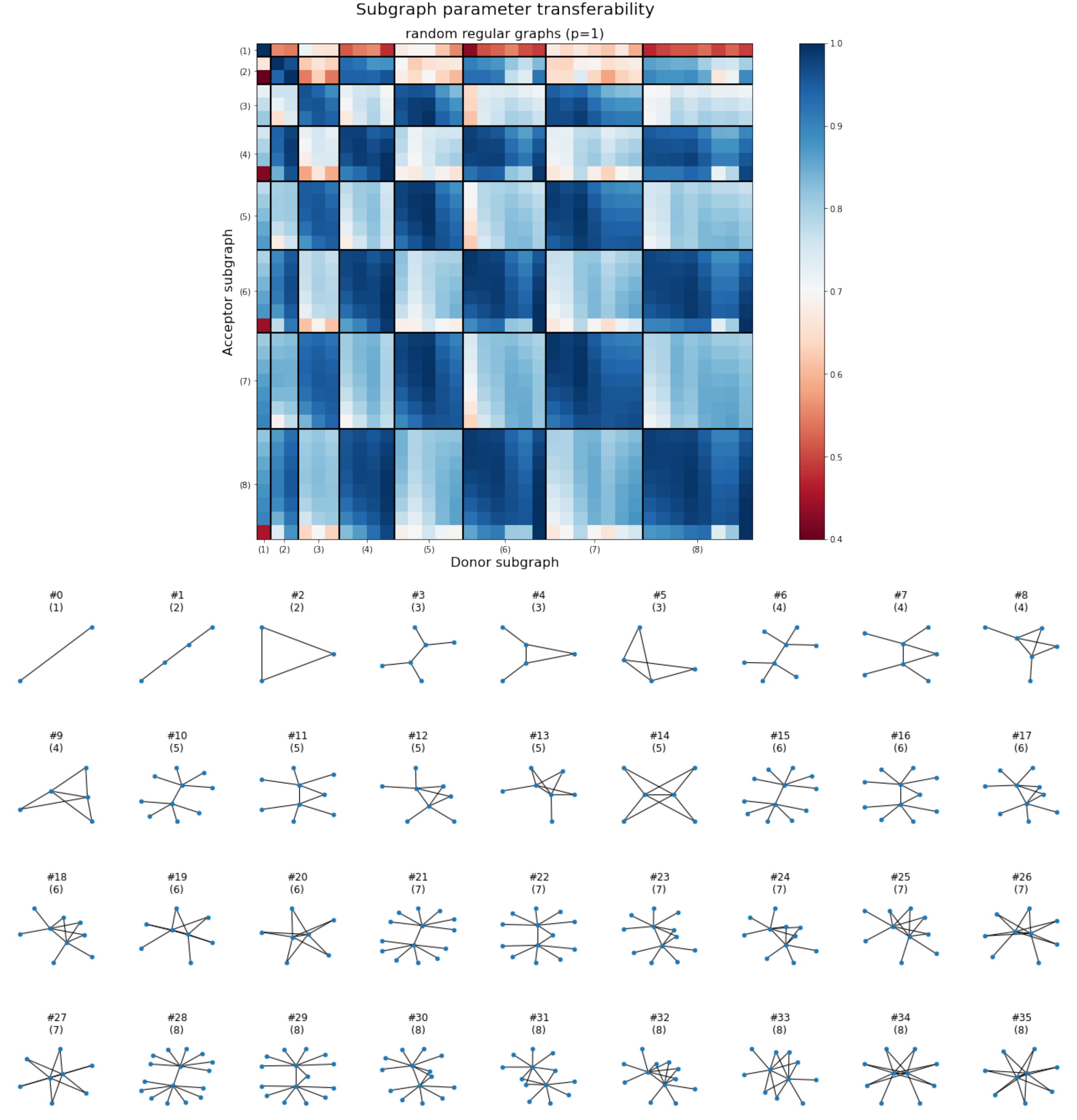}
    \caption{Transferability map between all subgraphs of random regular graphs with maximum node degree $d_\mathrm{max} = 8$, for QAOA depth $p = 1$. High (blue) and low (red) values represent good and bad transferability, respectively. Good transferability among even-regular and odd-regular random graphs and poor transferability across even- and odd-regular graphs in both directions are observed.}
    \label{fig:heatmap1}
\end{figure*}

To further investigate transferability among regular graphs, we evaluate the subgraph transferability map between all possible subgraphs of $d$-regular graphs, $d \leq 8$; see Figure~\ref{fig:heatmap1}. The top panel shows the colormap of parameter transferability coefficients between all possible pairs of subgraphs of $d$-regular graphs ($d \leq 8$, 35 subgraphs total). Each axis is split into groups of $d$ subgraphs of $d$-regular graphs, and the color values in each cell represent the transferability coefficient \(T(D, A)\) computed for the corresponding directional pair of subgraphs \(D, A\), defined as follows. For every subgraph \(G\), we performed numerical optimization with 200 steps, repeated 20 times with random initial points. This process results in 20  sets of optimal parameters of the form \((\gamma_{G_i}, \beta_{G_i})\) pairs, the best of which we will denote as \((\gamma_{G*}, \beta_{G*})\). Doing so for the donor subgraph \(D\) and the acceptor subgraph \(A\), the transferability coefficient \(T(D,A)\) averages over the QAOA energy contribution of each $(\gamma_{D_i}, \beta_{D_i})$ on the acceptor subgraph \(A\) as follows:
\begin{equation}  \label{eq:transf_coeff}
    \text{T(D, A)} = \frac{1}{20}\sum_{i = 1}^{20}\frac{\text{A}(\gamma_{D_i},\beta_{D_i})}{\text{A}(\gamma_{A*},\beta_{A*})},
\end{equation}
where \(A(\gamma, \beta)\) is the QAOA MaxCut energy of subgraph \(A\) as a function of parameters \((\gamma, \beta)\).

Instead of averaging over the 20 optimal parameters of the donor subgraph, we could have considered only the contribution of the donor's best optimal parameters \((\gamma_{D*}, \beta_{D*})\) in the above equation. For most donors, however, these best parameters were universal and hence  yielded high transferability to most acceptors. However, in practice, because of a lack of iterations or multistarts, we may converge to non-universal optimal parameters, resulting in the donor's poor transferability with some acceptors. The likelihood of converging to these non-universal optima for random graphs is discussed in Supplement \ref{sec:suppA}. Universal and nonuniversal optimal parameters are discussed in detail in Section \ref{subsec:Parity}.   

This inconsistency was discussed for 3-regular and 4-regular subgraphs earlier in this section. For example, half of the local optima of 3 regular subgraphs have good transferability to 4-regular subgraphs while the half yield poor transferability, as shown in in Fig \ref{fig:landscapes}.  Thus, to reflect practical considerations and avoid such inconsistency, we average over the contributions of 20 optimal parameters of the donor subgraph in Equation (\ref{eq:transf_coeff}). It is worth noting here that this averaging over 20 optimal parameters can result in poor transferability, as seen for donor subgraph \#0 to acceptor subgraphs \#2, \#9, \#20, and \#35. For these cases, there is a considerable subset of the donor's optimal parameters that lead to poor transferability. All considered subgraphs are shown in the bottom panel of Figure~\ref{fig:heatmap1}. Note that parameter transferability is a directional property between (sub)graphs, and good transferability from (sub)graph D to (sub)graph A does not guarantee good transferability from A to D. This general fact can be easily understood by considering two graphs with commensurate energy landscapes, for which every energy maximum corresponding to graph D also falls onto the energy maximum for graph A, but some of the energy maxima for graph D do not coincide with those of graph A.

The regular pattern of alternating clusters of high- and low-transferability coefficients in Figure~\ref{fig:heatmap1} illustrates that the parameter transferability effect extends from 3-regular graphs to the entire family of odd-regular graphs, as well as to even-regular graphs, with poor transferability between the two classes. For example, the established result for 3-regular graphs is reflected at the intersection of columns and rows with the label ``(3)'' for both donor and acceptor subgraphs. The fact that all cells in the 3x3 block in Figure~\ref{fig:heatmap1}, corresponding to parameter transfer between subgraphs of 3-regular graphs, have high values, representing high mutual transferability, gives a good indication of optimal QAOA parameter transferability between arbitrary 3-regular graphs~\citep{brandao2018fixed}.

\subsection{General Random Graph Transferability}\label{subsec:general}

\begin{figure*}[h!]
    \centering
    \includegraphics[scale=0.32]{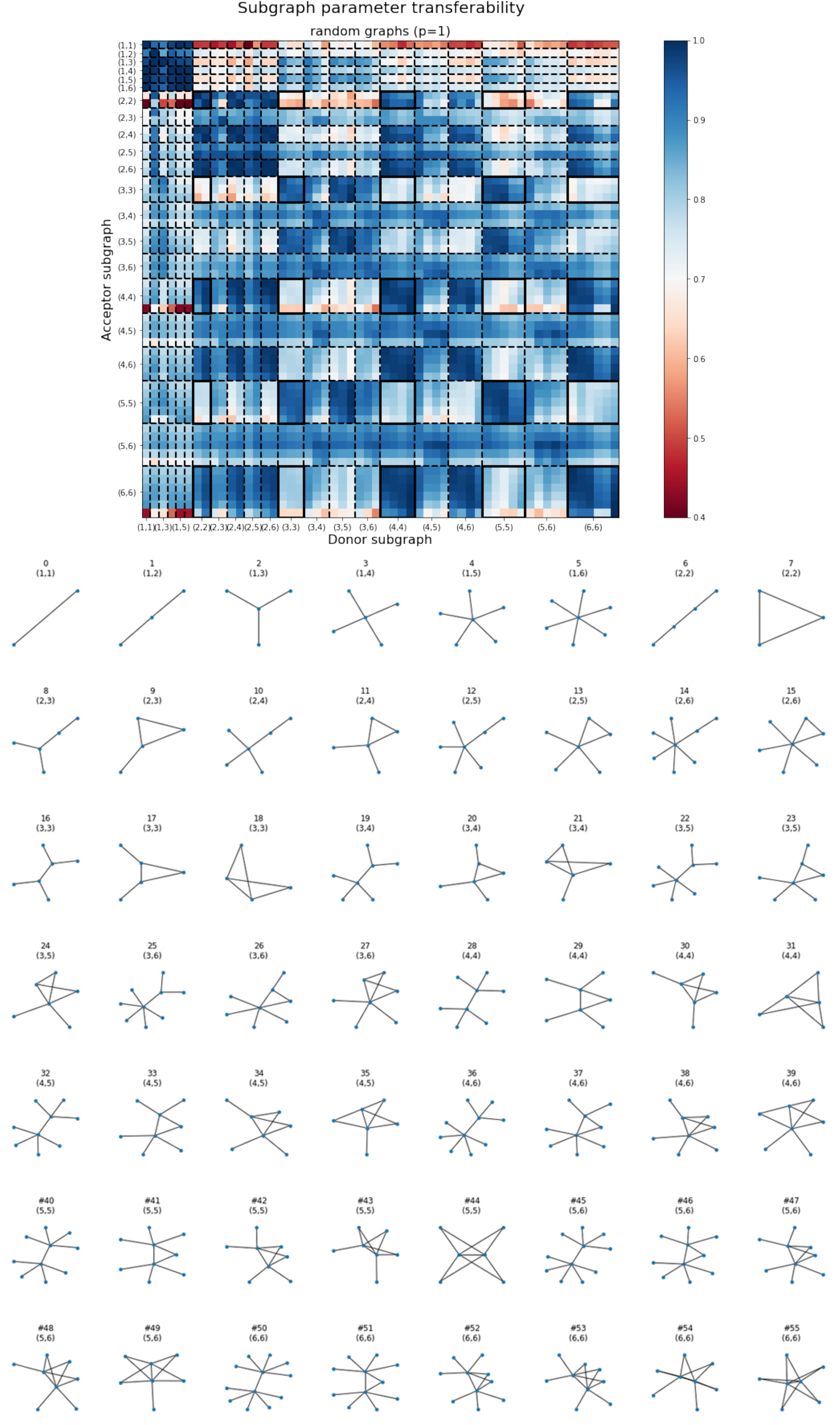}
    \caption{Transferability map between all subgraphs of random graphs with maximum node degree $d_\mathrm{max} = 6$, for QAOA depth $p = 1$. Subgraphs are visually separated by dashed lines into groups of subgraphs with the same degrees of the nodes forming the central edge. Solid black rectangles correspond to optimized parameter transferability between subgraphs of random regular graphs (Figure~\ref{fig:heatmap1}).
    }
    \label{fig:heatmap2}
\end{figure*}

Having considered optimal MaxCut QAOA parameter transferability between random regular graphs, we now focus on general random graphs. Subgraphs of an arbitrary random graph differ from subgraphs of random regular graphs in that the two nodes connected by the central edge can have a different number of connected edges, making the set of subgraphs of general random graphs much more diverse. The upper panel of Figure~\ref{fig:heatmap2} shows the transferability map between all possible subgraphs of random graphs with node degrees $d \leq 6$, a total of 56 subgraphs, presented in the lower panel. The transferability map can serve as a lookup table for determining whether optimized QAOA parameters are transferable between any two graphs.

Figure \ref{fig:heatmap2} reveals another important fact about parameter transferability between subgraphs of general random graphs. Subgraphs labeled as $(i, j)$, where $i$ and $j$ represent the degrees of the two central nodes of the subgraph, are in general transferable to any other subgraph $(k, l)$, provided that all $\{i, j, k, l\}$ are either odd or even. This result is a generalization of the transferability result for odd- and even-regular graphs described above.  Figure \ref{fig:heatmap2}, however,  shows that  a number of pairs of subgraphs with mixed degrees (not only even or odd) also transfer well to other mixed-degree subgraphs, for example, $\mathrm{subgraph}\,\#20\,(3, 4) \to \mathrm{subgraph}\,\#34\,(4, 5)$. The map of subgraph transferability provides a unique tool for identifying smaller donor subgraphs, the optimized QAOA parameters for which are also nearly optimal parameters for the original graph. The map can also be used to define the likelihood of parameter transferability between two graphs based on their subgraphs. As was the case for random regular graphs (see Figure \ref{fig:landscapes}), we see clustering of optimal parameters for subgraphs of random graphs in Figure \ref{fig:example}.

\begin{figure}[h!]
\centering
\includegraphics[scale=0.7]{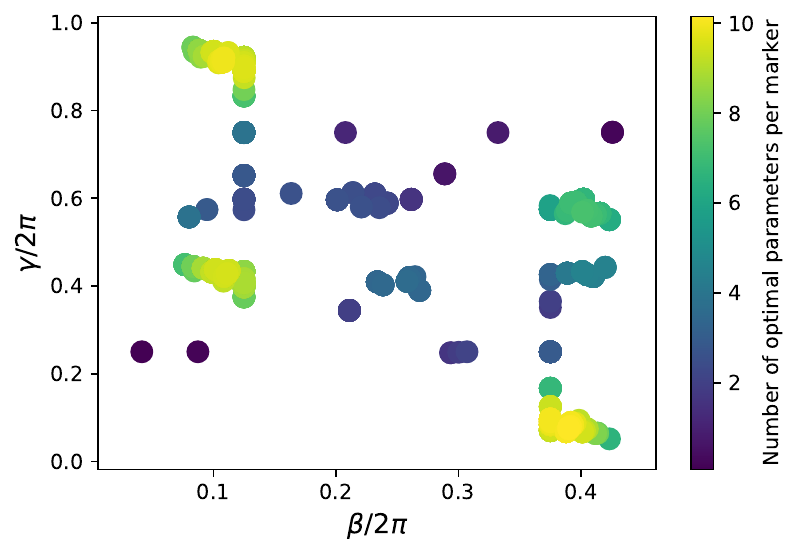}
\caption{\label{fig: subgraph_universal_param}
Distribution of optimal parameters of subgraphs with node degrees of central nodes ranging from 1 to 6 (total 56). Each subgraph was optimized with 20 multistarts, each of which is plotted in the figure.}
\end{figure}

\subsection{Parameter Transferability Examples}

\begin{figure}[h!]
    \centering
    \includegraphics[scale=0.12]{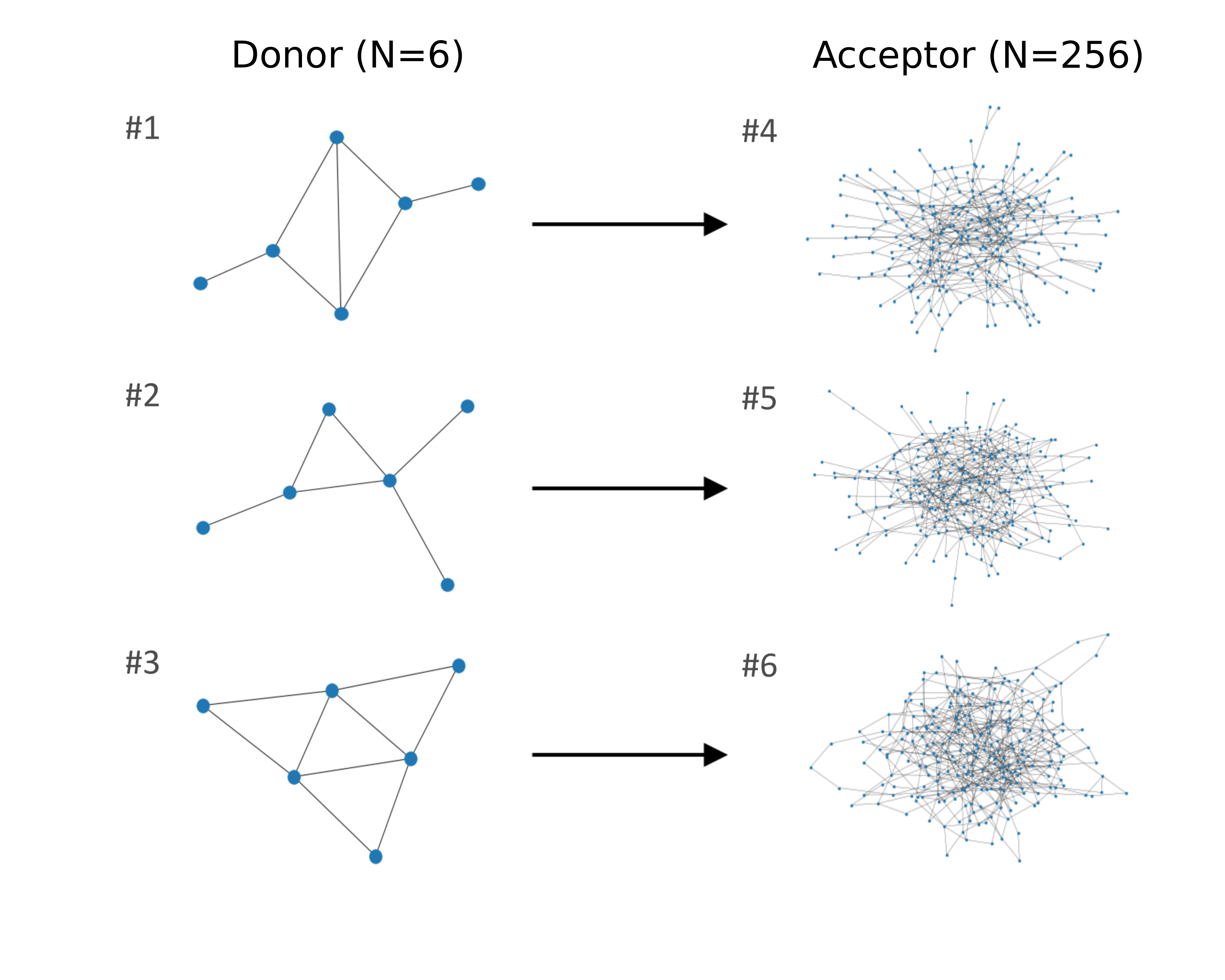}
    \caption{Demonstration of optimized parameter transferability between $N=6$ donor and $N=256$ acceptor random graphs. Using optimized parameters from the donor graph for the acceptor leads to the reduction in approximation ratio of 1.0\%, 2.6\%, and 1.0\% for the three examples (top to bottom, compared with optimizing the parameters for the acceptor graph directly, for $p = 1$).}
    \label{fig:example}
\end{figure}

We will now demonstrate that the parameter transferability map from Figure~{\ref{fig:heatmap2}} can be used to find small-$N$ donor graphs from which the optimized QAOA parameters can be successfully transferred to a MaxCut QAOA instance on a much larger acceptor graph. Initially, we consider three 256-node acceptor graphs to be solved and three 6-node donor graphs; see Figure~\ref{fig:example}. Table~\ref{tab:graphs} contains the details of the donor and acceptor graphs, including the total number of edges, their optimized QAOA energies, the energy of the optimal classical solution, and the approximation ratio. Graphs 1 and 4 consist exclusively of odd-degree nodes, graphs 2 and 5 contain roughly the same amount of both odd- and even-degree nodes, and graphs 3 and 6 contain exclusively even-degree nodes. The optimized QAOA parameters for the donor and acceptor graphs were found by performing numerical optimization with 20 restarts, and 200 iterations each. Additionally, we use a greedy ordering algorithm and an RMSprop optimizer, with a learning rate of 0.002. Table~\ref{tab:transfer} shows the results of the corresponding transfer of optimized parameters from the donor graphs \#\#1--3 to the acceptor graphs \#\#4--6, correspondingly. The approximation ratios obtained as a result of the parameter transfer in all three cases show only a 1--2\% decrease compared with those obtained by optimizing the QAOA parameters for the corresponding acceptor graphs directly. These examples demonstrate the power of the approach introduced in this paper.

\begin{table*}
\begin{center}
\caption{}
    \begin{tabular}{||c c c c c c||} 
    \hline
    Graph & Nodes & Edges & QAOA Energy & Energy (Opt) & Approx. Ratio\\ [0.5ex] 
    \hline\hline
    \#1 & 6 & 7 & 4.6481 & 6.0 & 0.7746\\ 
    \hline
    \#2 & 6 & 6 & 4.1272 & 5.0 & 0.8254\\
    \hline
    \#3 & 6 & 9 & 5.7050 & 6.0 & 0.9508\\
    \hline
    \#4 & 256 & 405 & 269.1192 & 363.0 & 0.7413\\
    \hline
    \#5 & 256 & 461 & 301.7699 & 400.0 & 0.7544\\
    \hline
    \#6 & 256 & 502 & 327.4132 & 430.0 & 0.7614\\ 
    \hline
    \end{tabular}
\\[10pt]
Details of donor and acceptor graphs, including number of nodes, number of edges, and both QAOA and classically optimized energies, along with their corresponding approximation ratios.
\label{tab:graphs}
\end{center}
\end{table*}

\begin{table}
\begin{center}
\caption{}
    \begin{tabular}{||c c c||} 
    \hline
    Transfer & QAOA Energy & Approx. Ratio\\ [0.5ex] 
    \hline\hline
    \#1 $\to$ \#4 & 226.2350 & 0.7334 (-1.0\%)\\ 
    \hline
    \#2 $\to$ \#5 & 293.8988 & 0.7347 (-2.6\%)\\
    \hline
    \#3 $\to$ \#6 & 323.8726 & 0.7753 (-1.0\%)\\
    \hline
    \end{tabular}
\\[10pt]
QAOA energies from transferred optimal parameters from 6-node donor graphs to 256-node acceptor graphs, along with their corresponding approximation ratios. The values in parenthesis show the reduction in the approximation ratio.
\label{tab:transfer}
\end{center}
\end{table}

\begin{figure*}
\centering
\includegraphics[width = \textwidth]{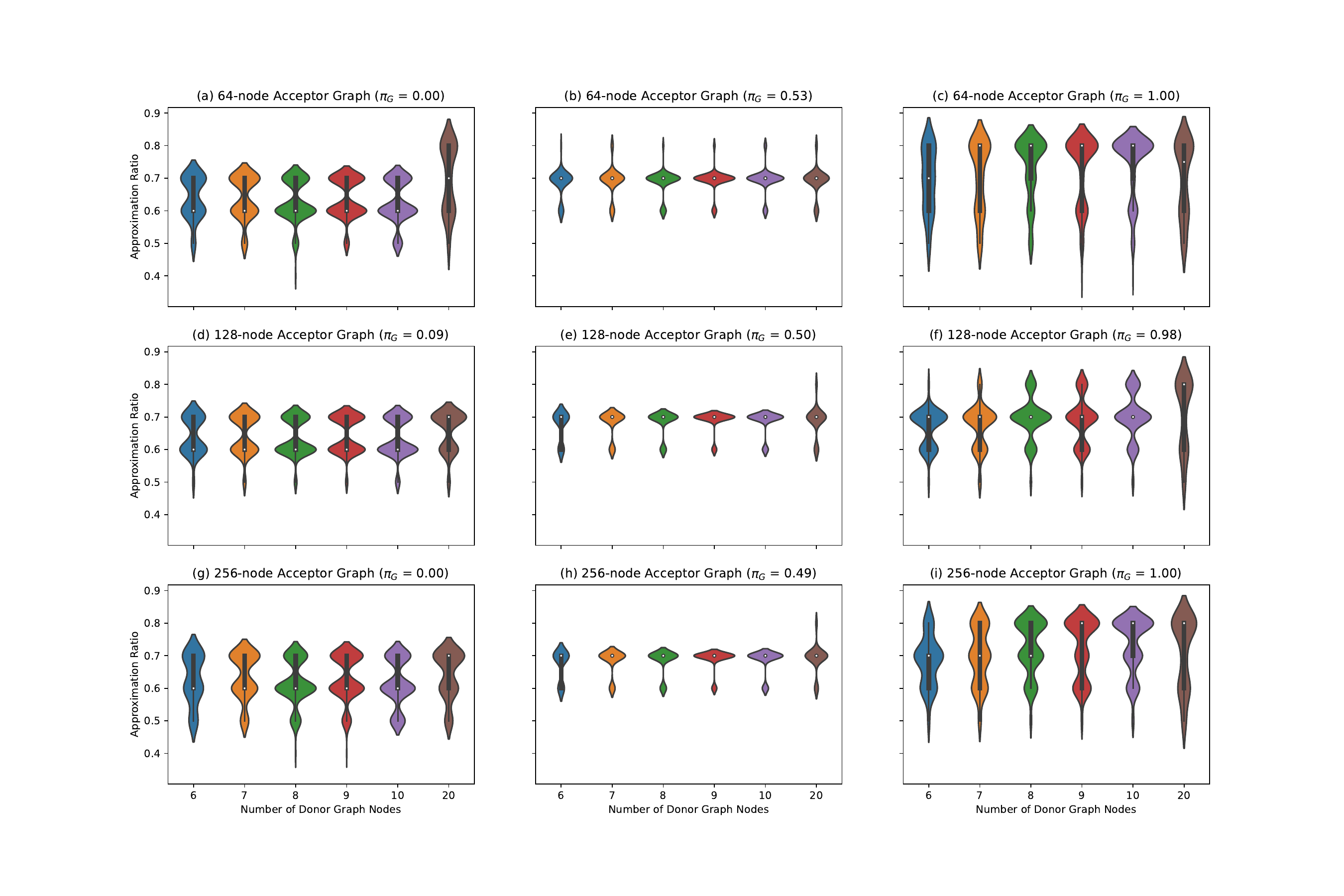}
\caption{\label{fig: violin plots} Approximation ratios for QAOA parameter transferability between lists of 6- to 20-node donor graphs and 64-node acceptor graphs (a)--(c), 128-node acceptors (d)--(f), and 256-node acceptors (g)--(i). The 64-node acceptors (top row) have the following parities: (a) 0.00, (b) 0.53, and (c) 1.00 eveness; 128-node acceptors (middle row) have the following parities: (d) 0.093, (e) 0.5, and (f) 0.98 evenness; and, 256-node acceptors (bottom row) have the following parities: (g) 0, (h) 0.49, and (i) 1.0 eveness.}
\end{figure*}

To extend our analysis of parameter transferability between QAOA instances, we perform transferability of optimal parameters between large sets of small donor graphs to a fixed, larger acceptor graph. In particular, we transfer optimal parameters from donors ranging from 6 to 20 nodes to 64-, 128-, and 256-node acceptor graphs. Figure \ref{fig: violin plots} shows the approximation ratio as we increase the number of donor graph nodes. These donor graphs were generated starting with graphs of exclusively odd-degree nodes and sequentially increasing the number of even-degree nodes until graphs of exclusively even-degree nodes were obtained. For each increasing number of node in a graph, ~100 donor graphs were generated and each of their 20 sets of optimal parameters (20 multistarts) were transferred to the acceptor graph. We see that there are a few cases for which we achieve an approximation ratio that is comparable to the native approximation ratio for each of the acceptor graphs. Most notably, we can achieve good transferability of optimal parameters to larger (i.e., 256-node) acceptor graphs without having to increase the size of our donor graph. Each row of Figure~\ref{fig: violin plots} corresponds to an increasing acceptor graph size, while each column corresponds to the parity of the acceptor graph (a formal definition and study of parity follow in the next section), with a transition from odd to even parity in graphs going from left to right. For the fully odd and fully even acceptor graphs, we notice a bimodal distribution in approximation ratios. Remarkably, for even acceptor cases, the bimodal distribution has one mode centered around the mean (white dot) and one above the mean. This points to the fact that, regardless of donor graph parity, one can achieve better parameter transferability when transferring optimal parameters to acceptor graphs with even parity. We see this transition from odd to even acceptor graphs in the way the bimodal distribution shifts, there being a monomodal distribution for the cases where the acceptor graphs are neither even nor odd.

The reason for this increased likeliness of good transferability to even acceptor graphs will be explored in future work. For now, we turn our focus to parity in graphs as an alternative metric for determining good transferability between donor-acceptor graph pairs, one that does not involve subgraph decomposition (and parameter transferability between individual subgraphs).

\subsection{Parity and Transferability} \label{subsec:Parity}

\begin{figure}
\centering
\includegraphics[scale=0.8]{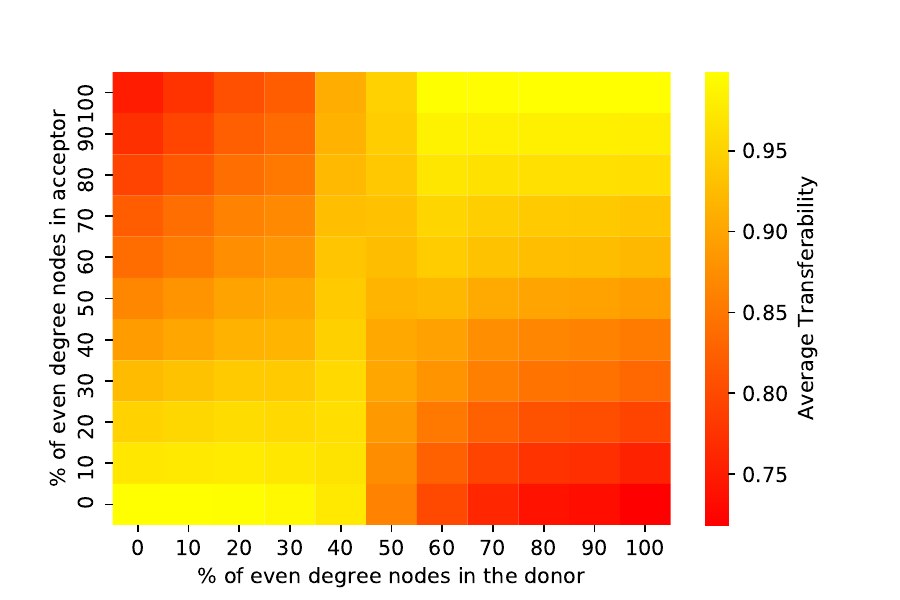}
\caption{\label{fig: parity_transferability}Transferability between 20-node random graphs as a function of the parity of degree of their vertices. The color of each block represents the average transferability of 100 graph pairs. As shown, graph pairs consisting of graphs of similar parity transfer well, while those of different parity transfer poorly.
}
\end{figure}

 As mentioned previously, the transferability maps of regular and random subgraphs suggest that the parity of graph pairs may affect their transferability. Here, we  define parity of a graph \(G = (V,E)\) to be the proportion of nodes of \(G\) with an even degree:  
    \begin{equation}\label{eq: parity}
        \pi_{G} = \frac{n_{even}}{|V|},
    \end{equation}
where $n_{even}$ is the number of even nodes in graph $G$. For this and upcoming sections, we focus on transferability between 20-node random graphs. That is, we perform optimal parameter transferability between 20-node donor and 20-node acceptor graphs. For every possible number of even-degree nodes \((0,2,4,\ldots, 20)\), we generated 10 graphs with distinct degree sequences, resulting in a total of 110 20-node random graphs, with maximum node degree restricted to 6.

The computed transferability coefficients among each graph pair, sorted by their parity, are shown in Figure~\ref{fig: parity_transferability}. Each block in the heatmap represents the average transferability coefficient of 100 graph pairs constructed from 10 distinct donor graphs and 10 distinct acceptor graphs. We can see that even graphs, those with $\pi_{G} = $ \(0.8-1\), and odd graphs, those with $\pi_{G} = $ \(0-0.2\), transfer well among themselves. However, the transferability between even donors and odd acceptors, as well as between odd donors and even acceptors, is poor.

This heatmap also suggests that the mutual transferability of a donor graph is not necessary for its good transferability with other random graphs, where
\emph{mutual transferability} of a graph \(G\) is a measure of how well the subgraphs of \(G\) transfer among themselves. Formally, it is defined as
\begin{align}
\begin{split}
    \text{MT}(G)
    &= {\sum_{d \in \{G\} }\sum_{a\neq d \in \{G\}} n_G(d) n_G(a)\frac{T(d, a)}{T_G}},\\
\end{split}
\end{align}
where \(\{G\}\) is the set of distinct subgraphs of graph \(G\), \(n_G(i)\) is the number of edges in \(G\) having subgraph \(i\), and 
$T_G = \sum_{d \in \{G\} }\sum_{a\neq d \in \{G\}} n_G(d) \cdot n_G(a)$ is the total number of subgraph pairs consisting of distinct subgraphs within \(G\). Graphs with low mutual transferability are those whose subgraphs transfer poorly among themselves. This is true for graphs with a nearly equal number of odd-parity and even-parity subgraphs since subgraph pairs of different parity report poor transferability coefficients, as shown in Figs.~\ref{fig:heatmap1} and \ref{fig:heatmap2}. In our case, such graphs are likely to be mixed-parity graphs, in other words, those with $\pi_{G} = $ \(0.4-0.6\). However, the results in Figure~\ref{fig: parity_transferability} show that these graphs have good transferability to nearly all random graphs in the data set.

\begin{figure*}
\centering
\includegraphics[width = \textwidth]{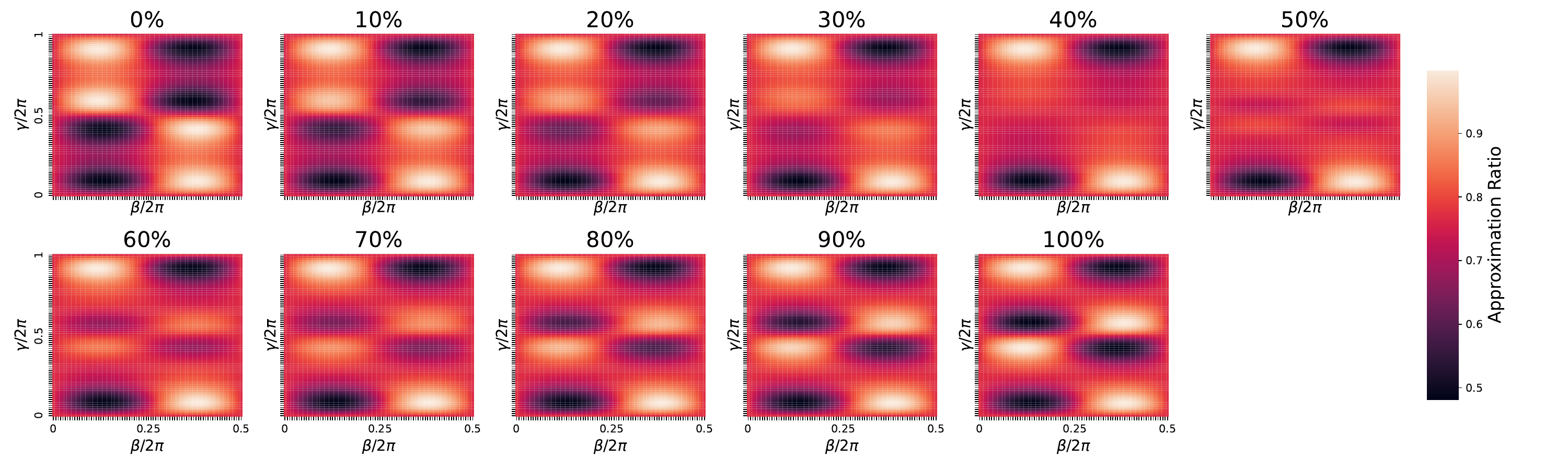}
\caption{\label{fig: centers_progression}Energy landscapes of some 20-node graphs sorted by parity. Each subplot is the average energy landscape of 10 20-node random graphs with the specified parity.
}
\end{figure*}

\begin{figure}
\centering
\includegraphics[scale=0.7]{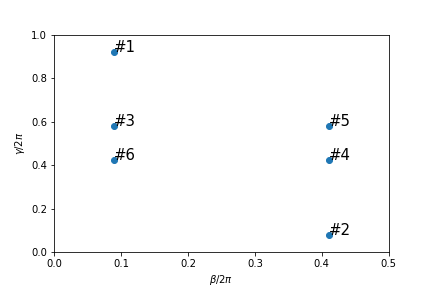}
\caption{\label{fig: centers_location} Energy landscapes of most 20-node random graphs had either local minima or maxima at one of these 6 centers. Here we label those points for later reference in the text.
}
\end{figure}

\begin{figure}
\centering
\includegraphics[scale=0.7]{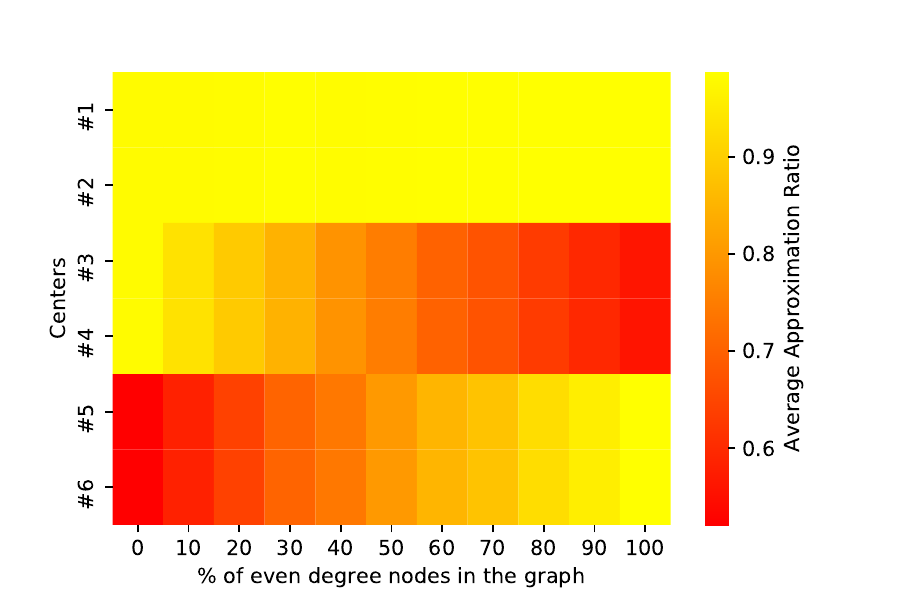}
\caption{\label{fig: centers_parity} Approximation ratios of the 110 20-node graphs at the 6 points in parameter space identified in Figure~\ref{fig: centers_location}. Parity, or the number of even-degree nodes in a graph, affects which centers correspond with minima and maxima. The first two centers, however, maximize every graph in the data set. 
}
\end{figure}

This trend can be explained by analyzing the energy landscapes of subgraphs. Most even- and odd-regular subgraphs have 4 maxima, two of which are universal for all regular subgraphs, as discussed in Section \ref{subsec:subgraphs}; the same trends were also observed in random subgraphs. Since energy landscapes of random graphs are sums of the energy landscape of its subgraphs, most 20-node random graphs share the same points, centers \(1,2\) in Figure~\ref{fig: centers_location}, as their local or global optima, as shown in Figure~\ref{fig: centers_parity}. On the other hand, the remaining two nonuniversal optimal parameters of regular subgraphs are shared only across regular subgraphs of similar parity. This property is also emergent in random graphs. In Figure~\ref{fig: centers_parity}, only odd random graphs share centers \(3,4\) as their local optima, while only even graphs share centers \(5,6\) as their local optima. However, mixed parity contained a nearly equal number of odd and even subgraphs. Since nonuniversal maxima of even subgraphs are minima for odd subgraphs and vice versa, these nonuniversal local optima blur on the energy landscapes of mixed-parity graphs. As a result, these graphs' landscapes  contain only universal maxima, as shown in the fourth energy landscape in Figure~\ref{fig: centers_progression}. With only universal parameters as their optimal parameters, mixed-parity graphs should indeed transfer well to all random graphs, as shown in the middle columns of Figure~\ref{fig: parity_transferability}.

The distribution of optimal parameters also explains poor transferability across random graphs of different parity. In Figure~\ref{fig: centers_parity}, the nonuniversal optimal parameters that maximize the MaxCut energy of odd random graphs, centers \(3,4\), also minimize that of even random graphs. Similarly, the nonuniversal optimal parameters that maximize the MaxCut energy of even random graphs, centers \(3,4\), also minimize that of odd random graphs. Consequently, transferring nonuniversal optimal parameters of even random graphs to odd random graphs and vice versa would result in poor approximation ratios, as evident in Figure~\ref{fig: parity_transferability}. 

Furthermore, above-average transferability for all graph pairs can be attributed to universal parameters. As shown in Figure~\ref{fig: similarity_metric}, all graph pairs have a true similarity or transferability coefficient greater than \(0.60\). Such a high lower bound can be attributed to universal parameters. Going back to Equation (\ref{eq:transf_coeff}), good transferability depends on whether the donor's optimal parameters \((\gamma_{D_i}, \beta_{D_i})\) optimize the acceptor graph as well. If most of the donor's optimal parameters are universal, in other words, are in the vicinity of centers \(1,2\) in Figure~\ref{fig: centers_location}, then the transferability coefficient will be high, regardless of the acceptor graph. In fact, Figure~\ref{fig: centers optimizer dist} in the Supplementary Material shows that, on average, all graphs reported at least half of their 20 optimal parameters as universal. As a result, they transfer well to other random graphs.

\subsection{Predicting transferability using subgraphs}

We have used the transferability coefficient to test whether an acceptor shares the same optimal parameters as its donor. In practice, this quantity is unknown because it requires knowledge of the acceptor's maximum energy. In earlier examples, we used the parity of graphs to explain transferability among random graphs, but the parity of a graph is just one emergent property from its subgraphs. Using subgraphs directly, we devise a subgraph similarity metric \(SS\) to predict the transferability ratio between a donor graph \(D = (V_D, E_D)\) and an acceptor graph \(A = (V_A, E_A)\) as follows: 
    
\begin{equation}\label{eq:similarity metric}
    \text{SS}(D, A) = \sum_{d \in \{D\} }\sum_{a \in \{A\}} n_D(d) n_A(a)\frac{\text{T}(d, a)}{|E_D|\cdot|E_A|},
\end{equation}
where \(\{G\}\) is the set of distinct \(p = 1\) subgraphs of \(G\), \(n_G(g)\) is the number of edges in \(G\) that share the subgraph \(g\), and  \(|E_D|\cdot|E_A|\) is the total number of subgraph pairs across graphs \(D\) and \(A\). Hence, this similarity metric states that the transferability coefficient of a donor and acceptor is the average transferability coefficient of donor subgraph-acceptor subgraph pairs.  

\begin{figure}[h!]
\centering
\includegraphics[scale=0.22]{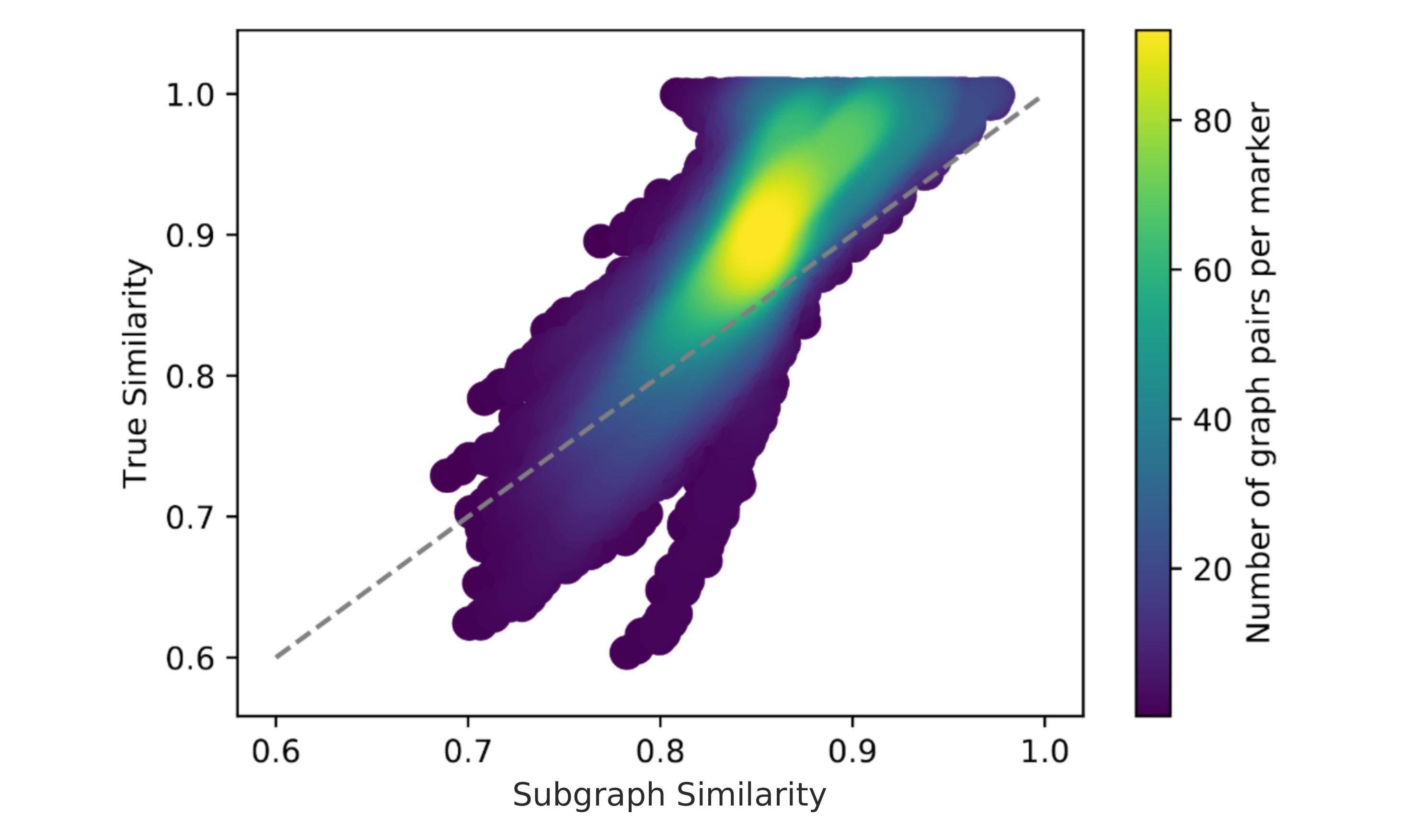}
\caption{\label{fig: similarity_metric}
Comparison of subgraph similarity metric \(SS\) with true similarity for \(110^2\) graph pairs consisting of 20-node graphs. The color indicates the density of points. For most graph pairs, \(SS\) underestimates the true similarity.}
\end{figure}

\begin{figure}[h!]
\centering
\includegraphics[scale=0.35]{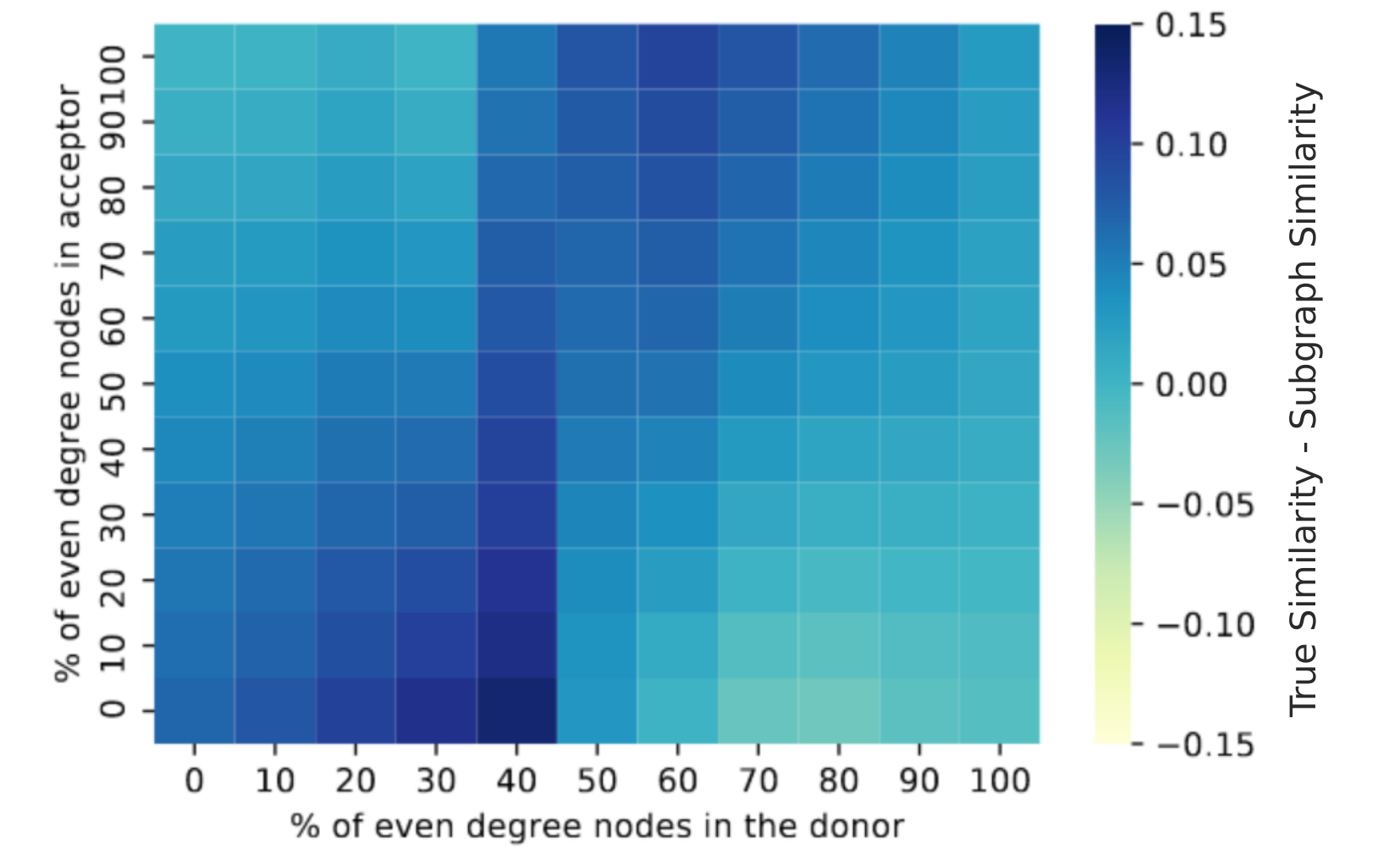}
\caption{\label{fig:subgraph sim diff} Differences between subgraph similarity metric \(SS\) and true similarity sorted by parity of donors and acceptors. \(SS\) largely underestimates true similarity for graph pairs consisting of graphs of the same parity.}
\end{figure}

In Figure~\ref{fig: similarity_metric} we compare this similarity metric with the true similarity or transferability coefficient. While this result does reveal a linear correlation between the two quantities,  the metric clearly under approximates the transferability coefficient by 0.05 units on average. In fact, Figure~\ref{fig:subgraph sim diff} shows that graph pairs with mixed-parity graphs as donors report the highest inconsistencies. This poor performance results from their constituent subgraphs. As discussed in Section \ref{subsec:Parity}, mixed-parity graphs consist of a nearly equal number of odd and even subgraphs. When optimized, these donor subgraphs may have nonuniversal optimal parameters. When transferred to an acceptor subgraph, the resulting transferability coefficient may be either poor or good, depending on the parity of that acceptor subgraph. While these nonuniversal optima do affect the subgraph similarity metric \(SS\), they do not affect true similarity. As shown in Figure~\ref{fig: centers_progression}, mixed-parity graphs' optimal parameters are universal. Thus, they transfer well to any random graph, regardless of its parity. Therefore, the subgraph similarity metric underestimates true similarity because it fails to capture that most optimal parameters of mixed-parity graphs are universal.  

\subsection{Predicting transferability using parity}

\begin{figure}[h]
\centering
\includegraphics[scale=0.4]{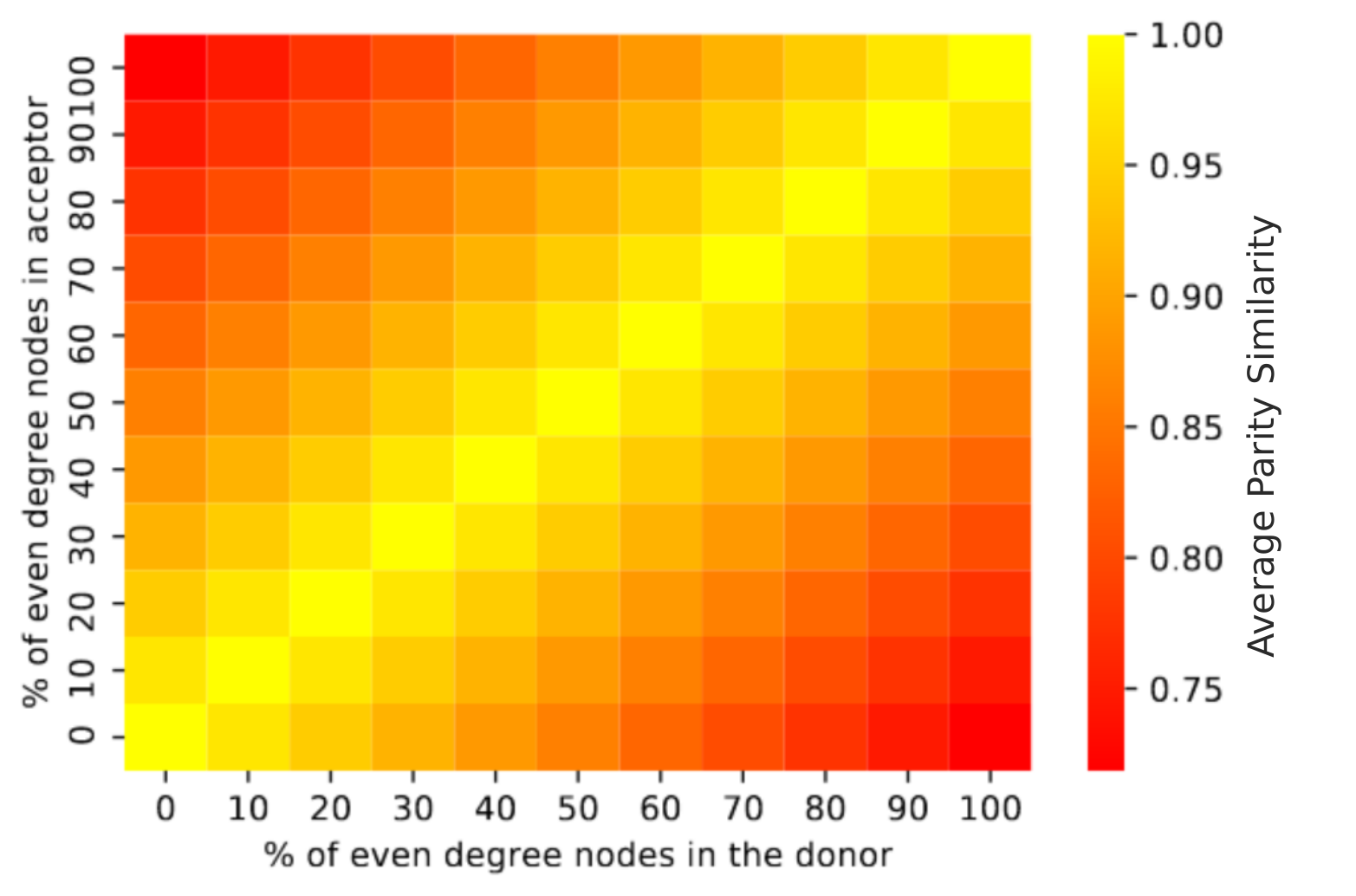}
\caption{\label{fig: parity plot SM2} Sorting of parity similarity metric \(PS\) for graph pairs based on the parity of the donor and the acceptor. Since \(PS\) assumes that graphs of similar parity transfer well, the diagonal reports the highest \(PS\).}
\end{figure}

Another approach to predicting transferability or similarity between two graphs is using their parity. In Section \ref{subsec:Parity} we observed that two graphs of similar parity have a high transferability ratio. If this correlation was ideal, then results shown in Figure~\ref{fig: parity_transferability} would resemble those in Figure~\ref{fig: parity plot SM2}. The parity similarity metric \(PS\) corresponding to the latter figure is  easy to compute:

\begin{equation} \label{eq:parity similarity metric 1}
    \text{PS}(D, A) = 1 -0.29\cdot|\text{Parity}(D) - \text{Parity}(A)|.
\end{equation}

\begin{figure}[h]
\centering
\includegraphics[scale=0.4]{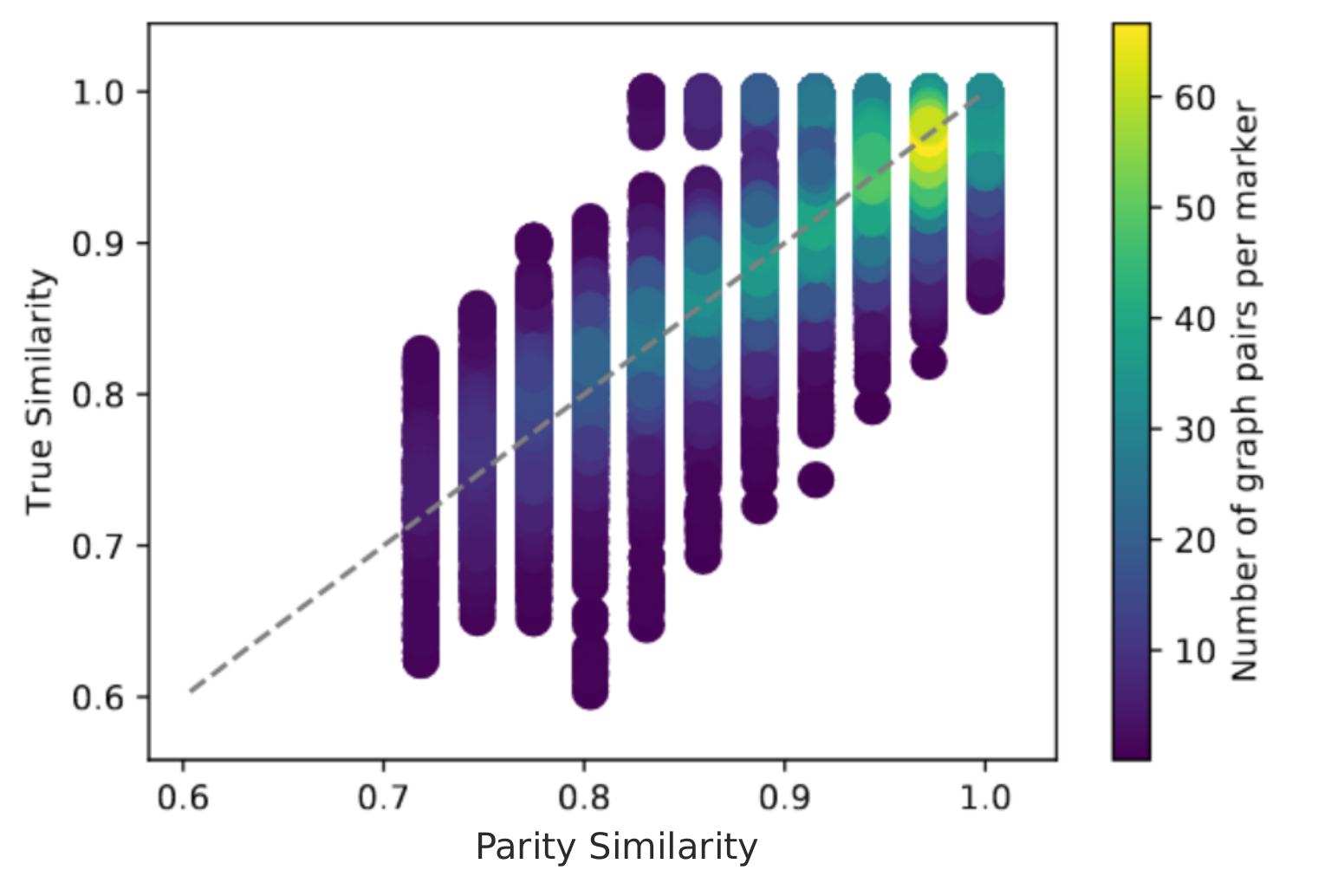}
\caption{\label{fig: similarity metric 2}Comparison of parity similarity metric \(PS\) with true similarity for \(110^2\) graph pairs consisting of 20-node graphs. The discrete columns occur because we cannot generate 20-node graphs with an arbitrary number of even-degree nodes.}
\end{figure}

Thus, this metric penalizes graph pairs consisting of different parity graph pairs. Note that the lowest value of this metric is \(\approx 0.71\), which is consistent with results from Figure~\ref{fig: parity_transferability}. Figure \ref{fig: similarity metric 2} illustrates the performance of this new similarity metric. The plot contains discrete columns because it is not possible to generate 20-node graphs with an arbitrary number of even-degree nodes. In order to ensure that the sum of a degree sequence is even, the associated graphs vary only in even-degree nodes in increments of two, resulting in parity of \(0.0, 0.1, 0.2, \ldots 1.0\). This discretization is also reflected in the similarity metric.

\begin{figure}[h!]
\centering
\includegraphics[scale=0.35]{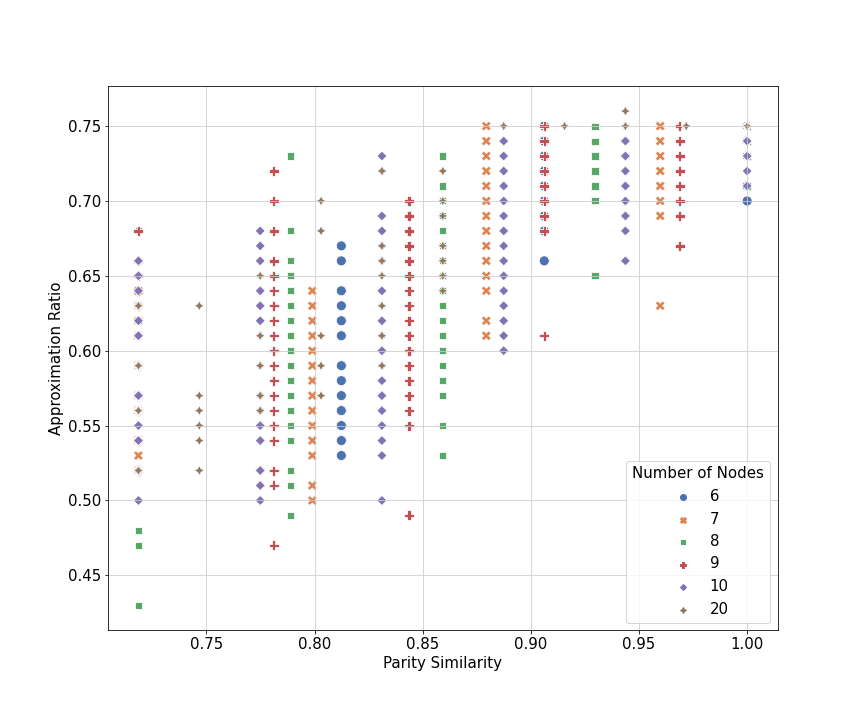}
\caption{\label{fig:approx v parity} For an increasing number of donor graph nodes, we see that parity can determine good transferability. For the case with subgraph transferability, we see that this does not depend on the number of nodes of the donor graph.}
\end{figure}

To test our similarity metric for the data set shown in \ref{fig: violin plots}, we compare our metric with the approximation ratio. Figure \ref{fig:approx v parity} shows that as parity between donor-acceptor pairs approaches 1 (i.e., the donor and acceptor graphs have the same parity), we achieve a higher approximation ratio. Noticeably, we see that we can have a good approximation ratio even if our parity similarity does not dictate so. This can be attributed to the fact that we are  exploiting only one structural feature from our graphs.

These results indicate that one can use a parity approach to determine good transferability between donor-acceptor pairs. Furthermore, one can generate a parity metric that  caters to specific graphs (please refer to Section \ref{sec:supp}).

\section{Conclusions and Outlook}\label{sec:conclusions}

Finding optimal QAOA parameters is a critical step in solving combinatorial optimization problems by using the QAOA approach. Several existing techniques to accelerate the parameter search are based on advanced optimization and machine learning strategies. In most works, however, various types of global optimizers are employed. Such a straightforward approach is highly inefficient for exploration because of the complex energy landscapes for hard optimization instances.

An alternative effective technique presented in this paper is based on two intuitive observations: (1) the energy landscapes of small subgraphs exhibit ``well-defined'' areas of extrema that are not anticipated to be an obstacle for optimization solvers (see Figure~\ref{fig:landscapes}), and (2) structurally different subgraphs may have similar energy landscapes and optimal parameters. A combination of these observations is important because, in the QAOA approach, the cost is calculated by summing the contributions at the subgraph level, where the size of a subgraph depends on the circuit depth $p$.

With this in mind, the overarching idea of our approach is solving the QAOA parameterization problem for large graphs by optimizing parameterization for much smaller graphs and reusing it. We started with studying the transferability of parameters between all subgraphs of random graphs with a maximum degree of 8. Good transferability of parameters was observed among even-regular and odd-regular subgraphs. At the same time, poor transferability was detected between even- and odd-regular pairs of graphs in both directions, as shown in Figs.~\ref{fig:heatmap1} and~\ref{fig:heatmap2}. This experimentally confirms the proposed approach.

A remarkable demonstration of random graphs that generalizes the proposed approach is the transferability of the parameters from 6-node random graphs (at the subgraph level) to 64-node random graphs, as shown in Figure~\ref{fig:example}. The approximation ratio loss of only 1--2\% was observed in all three cases. Furthermore, we demonstrated that one need not increase the size of the donor graph to achieve high transferability, even for acceptor graphs with 256 nodes.

Following the subgraph decomposition approach, we showed that one can determine good transferability between donor-acceptor graph pairs by exploiting their similarity based on parity. We see  a good correlation between subgraph similarity and parity similarity. In the future, we wish to address the exploitation of graph structure to determine good donor candidates, since subgraph similarities involve overhead calculations of QAOA energies for each pair of donor-acceptor subgraphs.

One may notice that we studied parameter transferability only for $p = 1$, where the subgraphs are small and transferability is straightforward. However, our preliminary work suggests that this technique will also work for larger $p$, which will require advanced subgraph exploration algorithms and will be addressed in our following work. In particular, we wish to explore the idea of generating a large database of donor graphs and, together with a graph-embedding technique, obtain optimal QAOA parameters for transferability. We hope that by training a good graph-embedding model, we will be able to apply our technique to various sets of graphs and extend our approach to larger depths. A machine learning approach has been used to determine optimal QAOA parameters \citep{khairy2020learning}, but a study of machine learning for donor graph determination is still an open question.

Another future direction is to determine whether the effects of parity of a graph hold for \(p>1\). In particular, we found that the parity of a graph affects the distribution of optimal parameters, as shown in Figs.~\ref{fig: centers_parity} and \ref{fig: centers optimizer dist}. It remains to be seen whether parameters concentrate for \(p>1\) and, if so, how parity affects their distribution. Analysis of these trends will be critical for the applicability of \(PS\) for \(p>1\).

This work was enabled by the very fast and efficient tensor network simulator QTensor developed at Argonne National Laboratory~\citep{qtensor}. Unlike state vector simulators, QTensor can perform energy calculations for most instances with $p \leq 3$, $d \leq 6$ and graphs with $N \sim\!1,000$ nodes very quickly, usually within seconds. For this work we computed QAOA energy for 64-node graphs with $d \leq 5$ at $p = 1$, a calculation that took a fraction of a second per each execution on a personal computer. With state vector simulators, however, even such calculations would not have been possible because  of the prohibitive memory requirements for storing the state vector.

As a result of this work, finding optimized parameters for some QAOA instances will become quick and efficient, removing this major bottleneck in the QAOA approach and potentially removing the optimization step altogether in some cases, eliminating the variational nature of QAOA. Moreover, our approach will allow finding parameters quickly and efficiently for very large graphs for which it will not be possible to use simulators or other techniques.

Our method has important implications for implementing QAOA on relatively slow quantum devices, such as neutral atoms and trapped-ion hardware, for which finding optimal parameters may take a prohibitively long time. Thus, quantum devices will be used only to sample from the output QAOA state to get the final solution to the combinatorial optimization problem. Our work will ultimately bring QAOA one step closer to the realization of quantum advantage.

\section*{Acknowledgments}

This research was developed with funding from the Defense Advanced Research Projects Agency (DARPA). The views, opinions and/or findings expressed are those of the author and should not be interpreted as representing the official views or policies of the Department of Defense or the U.S. Government. A.G., D.L. X.L., I.S., and Y.A. are supported in part by funding from the Defense Advanced Research Projects Agency. This work used in part the resources of the Argonne Leadership Computing Facility, which is a DOE Office of Science User Facility supported under Contract DE-AC02-06CH11357. The authors  thank Ruslan Shaydulin for insightful discussions.

\bibliographystyle{unsrtnat}
\bibliography{references}

\section{Supplemental Material}\label{sec:supp}

\subsection{Predicting transferability using parity -- extension}\label{sec:suppA}

\begin{figure}[h]
\centering
\includegraphics[scale=0.6]{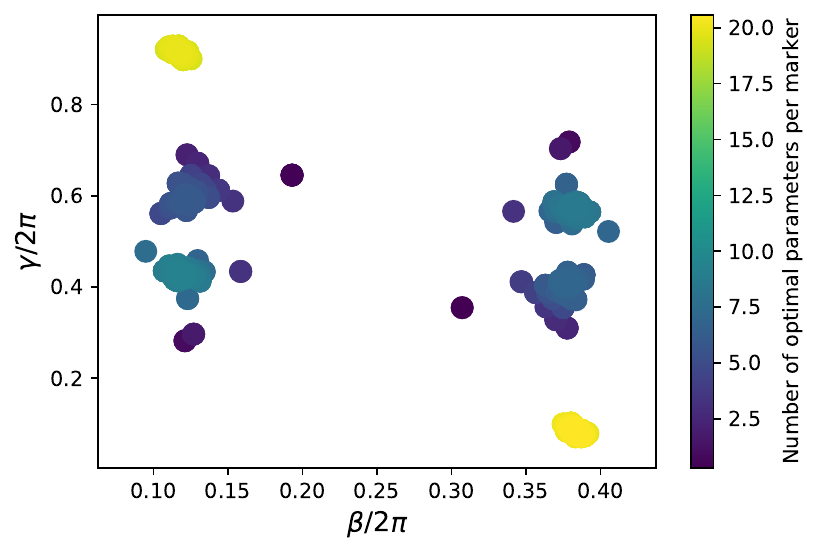}
\caption{\label{fig: centers graphs plot} Distribution of optimal parameters of 110 20-node random graphs. For each graph, we plot 20 optimal parameters, obtained from 20 multistarts with random initial points. }
\end{figure}

\begin{figure}[h]
\centering
\includegraphics[scale=0.6]{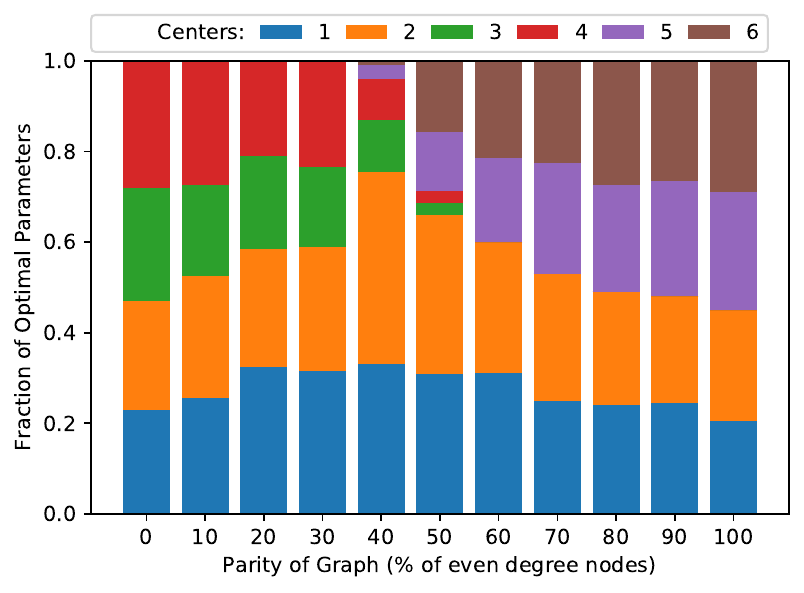}
\caption{\label{fig: centers optimizer dist} Sorting the computed optimal parameters of 110 20-node random graphs in Figure~\ref{fig: centers graphs plot} based on their vicinity to the 6 centers in Figure~\ref{fig: centers_location}. Clearly, for most graphs, at least half of the computer optimal parameters were universal, that  is,  near centers \(1,2\) .}
\end{figure}

Parity of graph correlates with the approximation ratios of a graph on the 6 centers shown in Figure~\ref{fig: centers_parity}. This trend can also be used to predict similarity, given two assumptions. First, the 20 computed optima are distributed among the 6 centers. This can be verified for 20-node graphs in Figure~\ref{fig: centers graphs plot}. Second, we can predict the location of these centers from the following observations from Figure~\ref{fig: centers_progression} and Figure~\ref{fig: centers optimizer dist}:
        \begin{itemize}
            \item All graphs have 4 local optima, two of which are universal. 
            
            \item For perfectly odd and perfectly even graphs, the local optima are distributed equally among the the four local optima.
            
            \item The fraction of optimal parameters which are universal increase and  that of optimal parameters which are nonuniversal decrease decrease as parity of a graph becomes mixed.
            
            \item Graphs that have \(c_3\) as their optima also have \(c_4\) as their optima. The same is true for centers \(c_5\) and \(c_6\).
        \end{itemize}
Algorithm \ref{alg:optima} combines these  observations to  predict the distribution of local optima for a graph among the 6 centers:
        \begin{algorithm}
    \caption{Optima Distribution}\label{alg:optima}
    \begin{algorithmic}[1]
    \Procedure{OptimaDistribution}{G}
        \State{\(n_{1,2} = n_{3,4} = n_{5,6} = 0\)}              
        \State{\(n_{1,2} = 10\)}
        \If{\(AR(G, c_3) >0.75\)}
        \State{\(n_{3, 4} = 10( (AR(G, c_3)-0.75 )/0.25)\)}
        
        \State{\(n_{1,2} += 10 -n_{3,4} \)}
        \ElsIf{\(AR(G, c_5) >0.75\)}
        \State{\(n_{5, 6} = 10( (AR(G, c_5)-0.75 )/0.25)\)}
        
        \State{\(n_{1,2} += 10 -n_{5,6} \)}
        
        \Else
        \State \(n_{1,2} += 10\)
        \EndIf
        
        \State{\textbf{return} \(n_{1,2}, n_{3,4}, n_{5,6}\)}
    \EndProcedure
    \end{algorithmic}
\end{algorithm}
where \(AR(G, c_i)\) is the approximation ratio of graph \(G\) at center \(c_i\) and \(n_{i,j}\) is the number of local optima distributed equally among centers \(c_i, c_j\).

Given these assumptions, for a given donor \(D\) and acceptor \(A\), we first compute OptimaDistribution\((D)\), that is, distribution of optimal parameters of the donor graph. Let \(n_i\) be the number of optima of the donor graph \(D\) occurring at center \(c_i\). Then the subgraph + parity similarity metric is 
    \begin{equation} \label{eq:parity similarity metric 2}
        \text{SPS}(D, A) = \frac{1}{20}\sum_{i = 1}^{6}n_i AR(A, c_i).
    \end{equation}

\subsection{Comparing metrics}

\begin{figure}[h]
\centering
\includegraphics[scale=0.3]{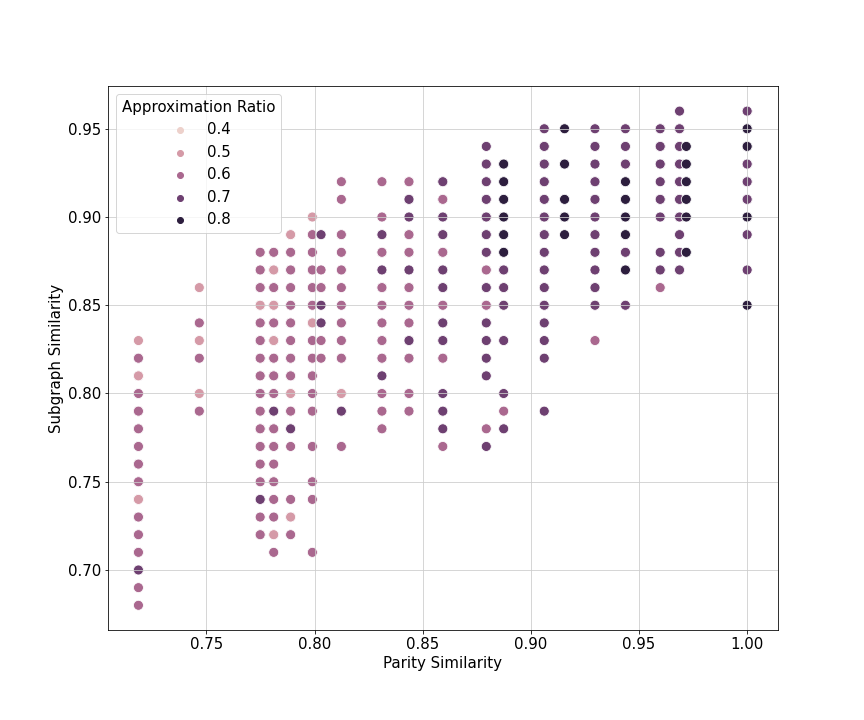}
\caption{\label{fig: subgraph v parity} Comparison of subgraph similarity with parity similarity. We see a correlation not only between similarity metrics but also with the approximation ratio.}
\end{figure}

To see correlation between subgraph similarity metric and parity similarity metric, we compare these metrics for the transferability studies performed on the large set of 6--20-node acceptor graphs and fixed 64-, 128-, and 256-node acceptor graphs. Figure \ref{fig: subgraph v parity} shows a correlation between subgraph similarity and parity similarity. Furthermore, we see that for either a high subgraph similarity or high parity similarity we obtain a good approximation ratio.

\begin{figure}[h]
\centering
\includegraphics[scale=0.4]{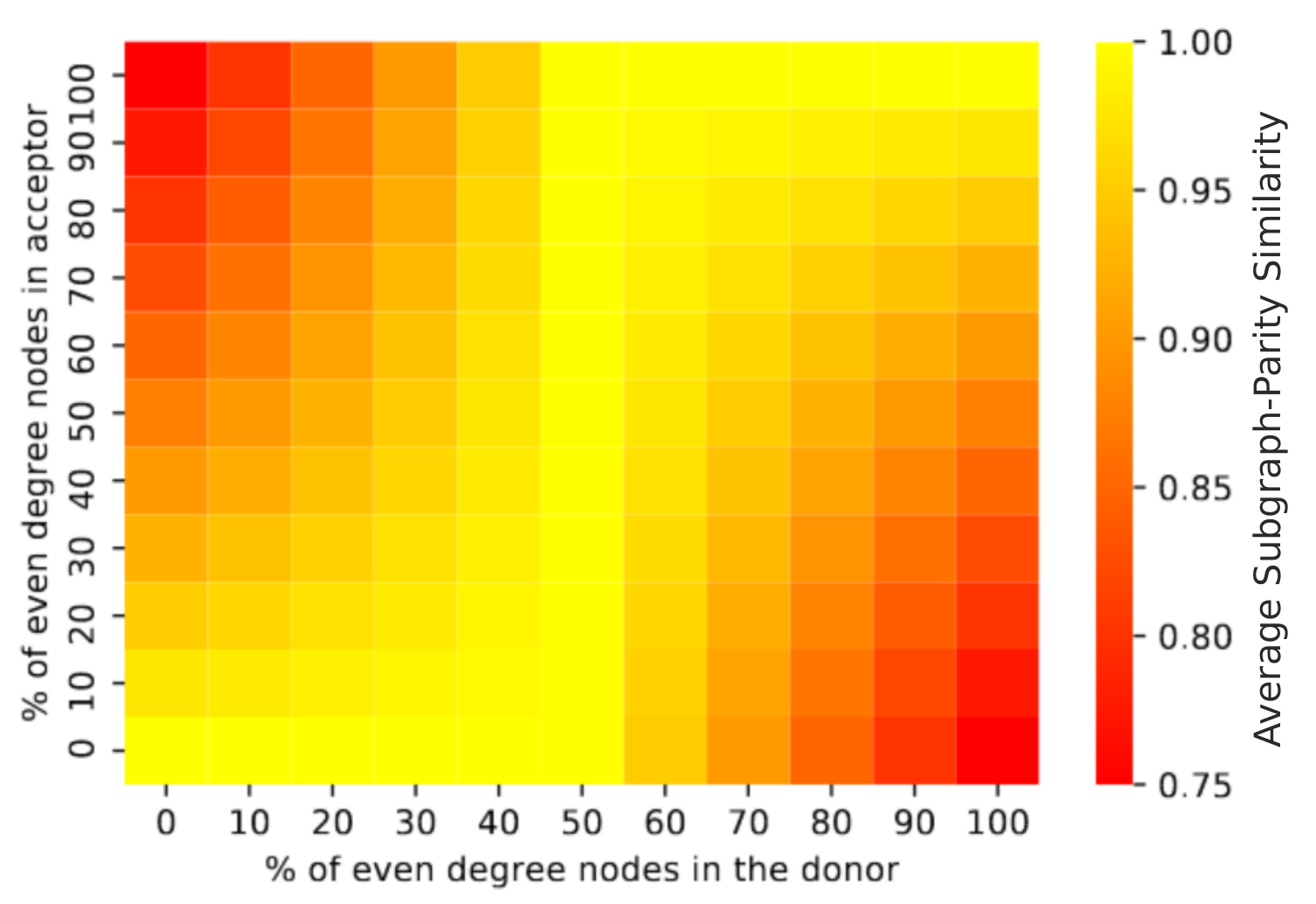}
\caption{\label{fig: parity plot SM3} Sorting parity similarity metric \(SPS\) for graph pairs based on the parity of the donor and the acceptor.}
\end{figure}

Furthermore, for the case of \(100^{2}\) 20-node graph pairs, we do a statistical comparison of the three metrics we propose. There results are given in Table \ref{tab:similarity}. While \(SPS\) may not have the best mean squared error or the best Pearson correlation coefficient, it best captures the relationship between parity and transferability coefficients. This is evident from  the heatmap in Figure~\ref{fig: parity plot SM3} closely resembling the heatmap of Figure~\ref{fig: parity_transferability}. However, the former assumes that the latter is symmetric about \(x=50\%\), which results in inconsistencies between \(SPS\) and true similarity.

\begin{figure}[h]
\centering
\includegraphics[scale=0.4]{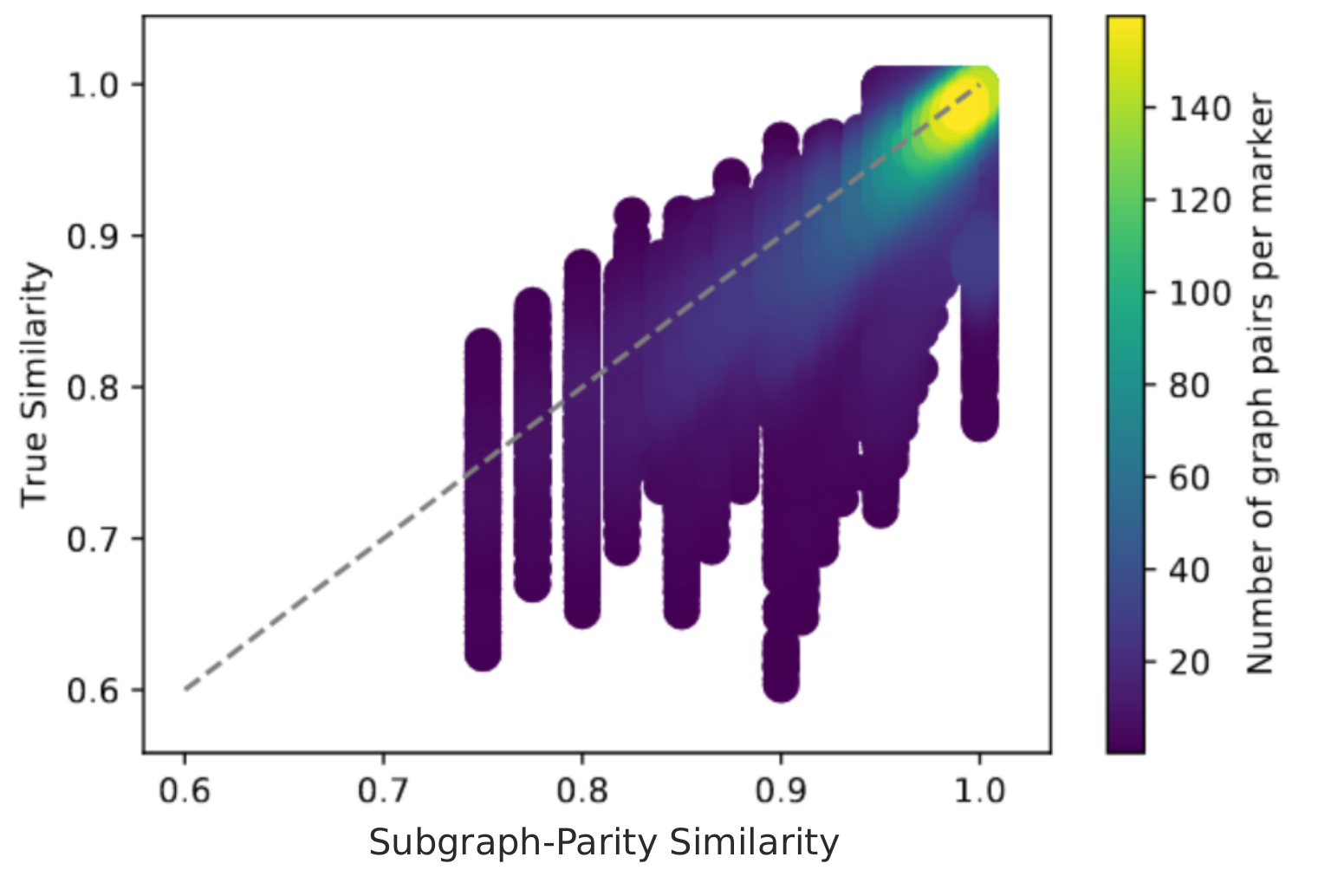}
\caption{\label{fig: parity similarity metric 2 plot} Comparison of subgraph similarity metric \(SPS\) with true similarity for \(110^2\) graph pairs consisting of 20-node graphs.}
\end{figure}

\begin{table}
\begin{center}
\caption{}\label{tab:similarity}
\begin{tabular}{||c c c c||} 
 \hline
 Statistical Comparison & SS & PS & SPS \\ [0.5ex] 
 \hline\hline
 Mean Squared Error & 0.0041 & 0.0025 & .0037 \\ 
 \hline
 Pearson Correlation Coefficient & 0.8298 & 0.7963 & 0.7721 \\
\hline
\end{tabular}
\end{center}
\end{table}

\end{document}